\documentclass[fleqn,10pt]{wlscirep}
\usepackage[utf8]{inputenc}
\usepackage[T1]{fontenc}
\usepackage{array}
\usepackage{amsmath}
\usepackage{chemfig}
\usepackage{amssymb}
\usepackage{hyperref}
\hypersetup{
    colorlinks=true,
    urlcolor= black,
    linkcolor = black,
    citecolor = black}
\usepackage{cite}
\usepackage{outlines}
\usepackage{nomencl}
\usepackage{graphicx}
\usepackage{float}
\usepackage{subcaption}
\usepackage{lipsum}
\usepackage{marvosym}

\title{Radical Pair Model for Magnetic Field Effects on NMDA Receptor Activity}

\author[1,\Letter]{Parvathy S Nair}
\author[2,3,4,\Letter]{Hadi Zadeh-Haghighi}
\author[2,3,4,\Letter]{Christoph Simon}

\affil[1]{Department of Physics, Indian Institute of Science Education and Research (IISER)
Tirupati, Tirupati-517507, Andhra Pradesh, India}
\affil[2]{Department of Physics and Astronomy, University of Calgary, Calgary, AB T2N 1N4, Canada.}
\affil[3]{Institute for Quantum Science and Technology, University of Calgary, Calgary, AB T2N 1N4, Canada.}
\affil[4]{Hotchkiss Brain Institute, University of Calgary, Calgary, AB T2N 1N4, Canada.}
\affil[\Letter]{email: \href{mailto:parvathysnair@students.iisertirupati.ac.in}{parvathysnair@students.iisertirupati.ac.in}; \href{mailto:hadi.zadehhaghighi@ucalgary.ca}{hadi.zadehhaghighi@ucalgary.ca}; \href{mailto:csimo@ucalgary.ca}{csimo@ucalgary.ca}}

\begin{abstract}
The N-methyl-D-aspartate receptor is a prominent player in brain development and functioning. Perturbations to its functioning through external stimuli like magnetic fields can potentially affect the brain in numerous ways. Various studies have shown that magnetic fields of varying strengths affect these receptors. We propose that the radical pair mechanism, a quantum mechanical process, could explain some of these field effects. Radicals of the form $[\mbox{RO}^\bullet \mbox{ Mg($\mbox{H}_2$O$)_n$}^{+\bullet}]$, where R is a protein residue that can be Serine or Tyrosine, are considered for this study. The variation in the singlet fractional yield of the radical pairs, as a function of magnetic field strength, is calculated to understand how the magnetic field affects the products of the radical pair reactions. Based on the results, the radical pair mechanism is a likely candidate for explaining the magnetic field effects observed on the receptor activity. The model predicts changes in the behaviour of the system as magnetic field strength is varied and also predicts certain isotope effects. The results further suggest that similar effects on radical pairs could be a plausible explanation for various magnetic field effects within the brain.
\end{abstract}
\begin{document}

\flushbottom
\maketitle

\thispagestyle{empty}

\section*{Introduction}

Magnetic field effects (MFEs) in biological systems have garnered considerable attention from the academic community. This has been studied in the context of numerous systems, such as magnetoreception\cite{Johnsen2005}, the working of the circadian clock\cite{Lewczuk2014}, genetics\cite{McCann1998}, and various others\cite{Ketchen1978,Fan2021,ZadehHaghighi2022,Buchachenko2014,Albuquerque2016,McKay2007,Yang2023,Adair2000,Bingi2003,Maffei2014,Karimi2020,2018}. In the last few decades, there has been significant progress in modelling different MFEs in these systems based on quantum mechanical phenomena\cite{Kim2021}. One model that has been quite promising in explaining some of these effects is the radical pair mechanism (RPM)\cite{Timmel1998}. This mechanism explains the behavior of radical pair electrons, which are pairs of correlated electrons formed on separate molecules from the rupture of chemical bonds or through electron hopping. The RPM model has been most commonly applied in explaining avian magnetoreception in migratory birds\cite{Hore2016}. It has also been used to model magnetoreception in other migratory organisms\cite{Mouritsen2018}, as well. 

There are numerous phenomena involving MFEs in biological systems that have yet to be explained by any mechanisms. One of the most fascinating areas to study such phenomena is the brain. It has been found that exposure to magnetic fields affects various functions in the brain. The frequency of neuron firing, pain sensitivity and inhibition, production of melatonin, and functioning of the pineal gland are some examples of these effects\cite{ZadehHaghighi2022}. 

Understanding the effects of magnetic fields on the brain is especially important since magnetic stimulations are already a well-known and well-used method for non-invasive brain stimulation used as both diagnostic and therapeutic tools\cite{Simons2005,Antczak2021,MAntczak2021}. Since magnetic stimulations can modulate various brain functions, they also have the potential to be used as a treatment for numerous neuropsychiatric diseases\cite{Luber2013,NOJIMA2020}. Understanding the mechanism by which magnetic fields affect important parts of the brain could potentially lead to further advances in such non-invasive treatment tools. 

Despite the existence of a large number of studies on these phenomena, there still needs to be more understanding of the mechanisms by which most of these effects occur. However, the RPM model has recently been applied to studying various phenomena within the brain with promising results. According to this mechanism, the electrons on the RP involved evolves under the influence of various interactions, like the Zeeman and hyperfine interactions, into a superposition of singlet and triplet states. The rate and yield of the final products formed from these states are affected by external magnetic fields, which is the cause of the various experimental observations. Lithium effects on hyperactivity\cite{ZadehHaghighi2021}, magnetic field and lithium effects on the working of the circadian clock\cite{ZHaghighi2022}, hypomagnetic field effects on neurogenesis\cite{Rishabh2022}, hypomagnetic field effects on microtubule reorganization\cite{ZH2022}, and Xenon-induced general anesthesia\cite{Smith2021} are examples of the various phenomena that have been modelled using this mechanism. These results encourage us to apply the mechanism to understand other phenomena within the brain.

One of the examples of such phenomena is the MFEs on the functioning of the N-Methyl-D-Aspartate receptors (NMDARs) within the brain. Several experimental studies successfully show that both static and oscillating magnetic fields of different strenghts can affect the activity of NMDARs\cite{zgn2019,Salunke2013}. These receptors are critical in all stages of development in higher organisms and are heavily involved in various brain functions, such as neuronal development and synaptic plasticity\cite{Paoletti2013}. The wide range of roles this receptor plays in the brain makes it a rich subject for research, with numerous studies dedicated to understanding its structure, function, and physiological roles over the years\cite{SanzClemente2012,PAOLETTI2007}. The variations in the different subunits that form this receptor increase the complexity and functionality of the various forms of the receptor\cite{Monyer1992,Furukawa2005}. As these receptors are deeply involved in the development and functioning of the brain, they are also crucial players in numerous neurological and psychological disorders. NMDARs have been shown to be involved in ischemic strokes, schizophrenia, excitotoxic brain injury, memory and learning impairments, and various other disorders\cite{Hansen2017,Liu2019,Nakazawa2020,Zhou2013}, thus making it extremely important that we understand the effects of external MFEs on these receptors and how they affect different processes within the brain. 

Hirai et al. demonstrated in two separate studies\cite{Hirai2003,Hirai2005} that immature rat hippocampal cells cultured in the presence of a 100 mT sustained or repetitive static magnetic field exhibit an increase in intracellular free $\mbox{Ca}^{2+}$ ion concentration upon increasing NMDA concentration. In both experiments, upregulation of certain mRNAs that encode some of the NMDAR subunits was observed.

We propose that an RPM model could provide an explanation for the observed effects. The MFE observed in these experiments may be occurring at any level, ranging from the transcription of the NMDAR subunits mRNAs to a direct effect on the functioning of the receptor. We attempt to identify the possible radical pairs involved in the various reaction pathways in these levels that may be affected by the static magnetic field (SMF). The potential RPs are identified based on the experimental evidence presented previously and an understanding of the mechanisms involved in the formation and function of the NMDAR. We study the effects of the magnetic field by calculating the product yield of the possible radical pair reactions involved in the system. On the basis of our findings, we conclude that an RPM model could provide a plausible explanation for the magnetic field effects observed on the NMDAR activity. This also provides further indication that radical pairs could be a key player in various magnetic field effects that occur in the brain. We further predict variations in the behaviour of the system as the magnetic field strength varies, which could be experimentally studied. We also show that there may be certain isotope dependent effects that could also be of experimental interest. 

\section*{Results}

\subsection*{Results from Previous Experiments}
Hirai et al.\cite{Hirai2003} conducted an experiment in which immature rat hippocampal cells were cultured for three days in vitro (DIV) in the presence of 100 mT SMF. Following the exposure, an increase in the intake of $\mbox{Ca}^{2+}$ ions was observed as the NMDA concentration within the cells was increased. In addition, an upregulation in the expression of mRNA encoding the NR1, NR2A-D, and NR3A subunits of NMDAR was observed in these cells following exposure.

\begin{table}[ht]
\centering
\begin{tabular}{ | m{2.7cm} | m{3cm}| m{3cm} | m{3cm} |} 
\hline
NMDA Concentration($\mu$M) & Cytoplasmic $\mbox{Ca}^{2+}$ Conc. in Control Setup(\%) & Cytoplasmic $\mbox{Ca}^{2+}$ Conc. at 100 mT(\%) & Ratio of Conc. at 100 mT vs. Control Setup\\
\hline
1 & 6.2 & 11.5 & 1.8 \\
\hline
3 & 15.1 & 23.7 & 1.6 \\
\hline
10 & 21.1 & 30.6 & 1.4 \\
\hline
30 & 24.7 & 33.2 & 1.3 \\
\hline
100 & 26.9 & 33.7 & 1.2 \\
\hline
\end{tabular}
\caption{\label{tab1}Approximate quantification of the experimental observations reported by Hirai et al.\cite{Hirai2003} of sustained exposure to SMF on the extent of $\mbox{Ca}^{2+}$ intake into rat hippocampal cells. Immature rat hippocampal cells were cultured for three days in vitro (DIV) in the presence of 100 mT SMF for this experiment. The values in the table are expressed as percentages over the maximal reading found at the end of the experiments from an addition of 10 $\mu$M of $\mbox{Ca}^{2+}$ ionophore(A23187).}
\end{table}

In a separate study conducted by Hirai et al.\cite{Hirai2005}, immature rat hippocampal cells were repeatedly exposed to 100 mT SMF for 15 minutes per day for eight days. Upon harvesting the cells 24 hours after the final exposure, it was found that there was once again a significant increase in intracellular $\mbox{Ca}^{2+}$ ion concentration. An upregulation in the mRNA expression of just the NR2B subunit of NMDAR by a factor (approximate) of 1.4 was also reported under the effect of such an exposure.

Tables \ref{tab1} and \ref{tab2} contain data from the experiments conducted by Hirai et. al. The data is approximate and has been extracted from plots in the two papers that visualize the variation of $\mbox{Ca}^{2+}$ ion concentration in the presence of the SMF, as NMDA concentration is varied.

\begin{table}[ht]
\centering
\begin{tabular}{ | m{2.7cm} | m{3.3cm}| m{3.3cm} | m{3cm} |} 
\hline
NMDA Concentration($\mu$M) & Cytoplasmic $\mbox{Ca}^{2+}$ in Control Setup(\%) & Cytoplasmic $\mbox{Ca}^{2+}$ at 100 mT(\%) & Ratio of $\mbox{Ca}^{2+}$ at 100 mT vs. Control Setup\\
\hline
1 & 3.9 & 6.2 & 1.6 \\
\hline
3 & 10.2 & 16.3 & 1.6 \\
\hline
10 & 17.9 & 25.8 & 1.4 \\
\hline
30 & 22.8 & 28.2 & 1.2 \\
\hline
100 & 25.1 & 28.4 & 1.1 \\
\hline
\end{tabular}
\caption{\label{tab2}Approximate quantification of the experimental observations reported by Hirai et al.\cite{Hirai2005} of repeated exposure to SMF on the extend of $\mbox{Ca}^{2+}$ intake into rat hippocampal cells. Immature rat hippocampal cells were repeatedly exposed to 100 mT SMF for 15 minutes per day for eight days for this experiment. The values in the table are expressed as percentages over the maximal reading found at the end of the experiments from an addition of 10 $\mu$M of $\mbox{Ca}^{2+}$ ionophore(A23187).}
\end{table}

From the same experiments, Hirai et al. also reported an increase in the binding activity of Activator Protein-1(AP-1), which is a transcription factor that has been shown to be an active regulator of NR2B transcription\cite{Qiang2005}.

\subsection*{Possible Role of Phosphorylation in MFE on NMDARs}

The first possible explanation for the MFE is an increase in NMDAR activity as a result of a rise in the number of mRNAs that form these receptors. Prybylowski et al.\cite{Prybylowski2002} demonstrated that an increase in the number of NR2 subunits within cerebellar granule cells causes an increase in the concentration of NMDARs within the cells. The same is most likely true for hippocampal cells.

As stated previously, an increase in the NR2B mRNA concentration is observed within the cells in the presence of an SMF\cite{Hirai2003,Hirai2005}. It is highly likely that an increase in mRNA concentration would also result in an increase in NR2B subunit concentration within the cells. According to the findings of Prybylowski et al., the increase in NR2B subunits should result in an increase in the number of NMDARs in the cell. An increase in the number of NMDARs provides more channels for the $\mbox{Ca}^{2+}$ ions to enter the cells, thus causing an increase in the concentration of these ions within the cells, as the concentration of NMDA in the culture increases. 

The upregulation in NR2B mRNA expression observed in the experiments could have occurred as the result of an MFE occurring at any stage of the mRNA formation. The increase in AP-1 binding activity reported by Hirai et al. could be a possible explanation for the same. AP-1 transcription factors typically consist of heterodimers of Jun and Fos proteins or homodimers of Jun proteins. The DNA binding activity of AP-1 is regulated by phosphorylation and dephosphorylation processes of the c-Jun protein, which belongs to the Jun protein family. Phosphorylation of Ser 63 and Ser 73 in the N-terminal domain of c-Jun stimulates the transcriptional activity of the protein\cite{Gupta1996, Kallunki1996}. Phosphorylation has been reported to be affected by MFEs\cite{Markov1997,Engstrm2002,Buchachenko2005}, making this process a likely candidate for the SMF effects.

As mentioned previously, it is also possible that the MFE occurs directly on the NMDAR or its subunits. Phosphorylation of serine and tyrosine sites in NMDAR subunits has been shown to regulate NMDAR activity\cite{Chen2007, Wang2014}. For instance, phosphorylation of Ser 896 and Ser 897 together increases NMDAR surface expression. Receptor currents increase in NMDARs containing NR2A subunits when Ser 1291 and Ser 1312 are phosphorylated. Tyrosine phosphorylation of NR2A also potentiates NMDAR currents. These phosphorylation processes can also be influenced by MFEs.

We propose that the MFEs reported on NMDAR activity may be modeled by an RPM model based on the effects of SMF on one of the aforementioned processes. The MFE could occur via either a Serine phosphorylation, which will be referred to as the "Ser Pathway," or a tyrosine phosphorylation, which will be referred to as the "Tyr Pathway."

\subsection*{RPM Model of Oxyradicals and Hydrated Magnesium Cations}

Buchachenko et al.\cite{LBuchachenko2014} theorized that the transfer of phosphate groups to proteins induced by protein kinases may involve an ion-radical mechanism. To confirm this, they studied the catalysis of phosphorylation of prothrombin by prothrombin kinases with $^{24}$Mg$^{2+}$ and $^{25}$Mg$^{2+}$ ions. They successfully demonstrated an isotope dependence in the kinase's efficiency, which could be due to the nuclear spin of the Mg$^{2+}$ isotopes. It was proposed that this indicates that the ion-radical mechanism could be a plausible mechanism for enzymatic phosphorylation processes. The mechanism involves the following reactions:
\[\mbox{RO}^{-} + \mbox{Mg(H}_2\mbox{O)}_n^{2+} \longrightarrow \mbox{RO}^\bullet + \mbox{ Mg($\mbox{H}_2$O$)_n$}^{+\bullet}\]
\[\mbox{RO}^\bullet + \chemfig{P(-[:65,0.75]O^{-})(-[:115, 0.75]^{-}O)(=[:270, 0.75]O)(-[:0, 0.75]ADP)} \longrightarrow \chemfig{P(-[:180,0.75]RO)(-[:65,0.75]O^{-})(-[:115, 0.75]^{-}O)(-[:270, 0.75]O^{\bullet})(-[:0, 0.75]ADP)} \longrightarrow \chemfig{P(-[:180,0.75]RO)(-[:65,0.75]O^{-})(-[:115, 0.75]^{-}O)(=[:270, 0.75]O)} + \mbox{ADP}^\bullet\]
 
According to this mechanism, phosphorylation could involve an oxyradical, $\mbox{RO}^\bullet$(where R represents a protein residue), and a hydrated magnesium cation radical, $\mbox{ Mg($\mbox{H}_2$O$)_n$}^{+\bullet}$. 

Depending on the pathway involved, the protein residue for our system could be either Serine or Tyrosine. Based on this, we propose that the MFEs can be modeled by an RPM model involving the RP $[\mbox{RO}^\bullet \mbox{ Mg($\mbox{H}_2$O$)_n$}^{+\bullet}]$ formed during this reaction. The reactions involved in the transcriptional level phosphorylation processes involve only the Ser Pathway, whereas the Ser or Tyr Pathways may be involved in the MFEs on NMDAR subunits. 

We model the MFEs on the RPs using a simplified Hamiltonian consisting only of Zeeman and hyperfine interactions (HFIs),which is a good approximation for sufficiently distant radicals\cite{Hore2016, Efimova2008}. We assume that radical electrons have the same g-value as free electrons, which is a good approximation at the field values we are interested in. We also consider only the isotropic Fermi contact contribution for the HFIs, as the molecular arrangement is likely to be random. Therefore, the Hamiltonian would have the following form:
\[\hat{H} = \omega(\hat{S}_{A_z} + \hat{S}_{B_z}) + a_{A}\hat{\boldsymbol{S}}_A\cdot \hat{\boldsymbol{I}}_{A} + a_{B}\hat{\boldsymbol{S}}_B\cdot \hat{\boldsymbol{I}}_B\]
Here, $\hat{\boldsymbol{S}}_A \mbox{ and } \hat{\boldsymbol{S}}_B$ are the electron spin operators of the RPs labelled $A$ and $B$, respectively. The Larmour frequency of the electrons due to Zeeman interaction is given by $\omega$. $\hat{\boldsymbol{I}}_A$ represents the nuclear spin operator associated with the oxyradical nucleus that contributes most to the HFI. 

We perform our calculations considering the contributions from the natural abundance of the different magnesium isotopes. It is well known that 10\% of naturally occurring magnesium is composed of the isotope $^{25}$Mg, which is a spinful isotope. In the equation above, $\hat{\boldsymbol{I}}_B$ represents the nuclear spin operator of any naturally occurring $^{25}$Mg nucleus present in the system. The terms $a_{A}$ and $a_B$ are the hyperfine coupling constants for each of the nuclei.

We use density functional theory (DFT) calculations to determine the hyperfine coupling constant (HFCC) values for all spinful nuclei involved. The details for the same have been given in the Methods section. For our calculations, we consider the HFI contribution from only the nucleus with the highest HFCC, which is a commonly used approximation. In the case of the Serine oxyradical, we find that the highest HF contribution is from a hydrogen nucleus with $a_{A} = 7.45\mbox{ mT}$. For the Tyrosine oxyradical, the highest HFCC is of one of the hydrogen nuclei with $a_{A_1} = 1.86\mbox{ mT}$. The HFCC of the $^{25}$Mg isotope is $a_B = -11.22$ mT. 

\subsection*{Singlet Yield Calculations}
The fractional singlet yield (FSY) of the reaction with and without the magnetic field indicates how the SMF affects the rate of formation of the different reaction products. This could lead to further changes in the reaction rates or product concentrations of various reactions involved in the pathways leading to the experimentally observed effects\cite{ZHaghighi2022}.

The FSY of the reaction is calculated by tracking the spin state of the RP during the reaction. This calculation can be carried out by solving the Liouville-von Neumann equation, which describes the evolution of the density matrix over time\cite{Timmel1998}. For a general singlet-born RP under the effect of a weak magnetic field, the fractional yield for periods larger than the lifetime of the RPs is found from the eigenvalues and eigenvectors of the Hamiltonian as:

\[\Phi_S = \frac{1}{M}\sum^{4M}_{m = 1}\sum^{4M}_{n = 1}|\langle m|\hat{P}^S|n\rangle|^2 \frac{k(k+r)}{(k+r)^2 + (\omega_m - \omega_n)^2} - \frac{k}{4(k+r)} + \frac{1}{4},\]

where $M = M_AM_B$, $M_X = \prod^{N_X}_i I_{iX}(I_{iX} + 1)$ is the nuclear spin multiplicity, $|m \rangle \mbox{ and } |n\rangle$ are the eigenstates of the hamiltonian, $\hat{H}$, with the eigenenergies given by $\omega_m \mbox{ and } \omega_n$, respectively, and $\hat{P}^S$ is the singlet projection operator acting on the electron spins. Here, we assume that the singlet and triplet reaction rates are equal and are denoted by $k$. Finally, $r$ is the relaxation rate or spin-coherence lifetime of the radical pairs\cite{Hore2019}.

As mentioned above, the FSY provides information about the yield of the different reaction products with and without the effect of the SMF. We connect this change in the yield of the products to the change in the intracellular concentration of $\mbox{Ca}^{2+}$ ions, in order to understand the MFE on the cells.

Due to the lack of experimental data for the exact values of the reaction rate, $k$, and relaxation rate, $r$, of the RP we consider, we explore the $k/r$ space for an approximate range of potential values. We consider the ratio of the FSY at the control magnetic field strength($B_0$) to the FSY at $100$\mbox{ mT}($B_{\mbox{exp}}$), which is the experimental field strength used:
\[S = \frac{\mbox{FSY at $B_0$}}{\mbox{FSY at $B_{\mbox{exp}}$}},\]

We plot this quantity against a range of possible $k$ and $r$ values. The magnitude of the ratio reflects the extent of the magnetic field effects on radical pair electrons, which we compare to the magnitude of the ratio of $\mbox{Ca}^{2+}$ concentrations within the cells, as shown in Table \ref{tab1} \& Table \ref{tab2}.

\subsubsection*{Ser Pathway}
\begin{figure}[h!]
\begin{subfigure}{.49\textwidth}
  \centering
  % include first image
\includegraphics[width=1\linewidth]{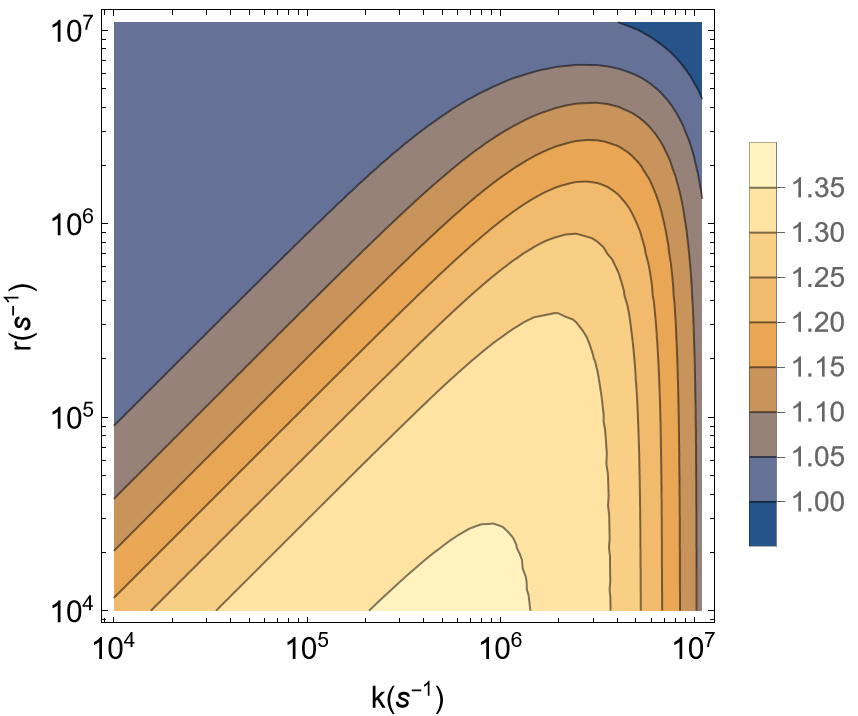}
  \caption{}
  \label{fig:1.1}
\end{subfigure}
\begin{subfigure}{.49\textwidth}
  \centering
  % include second image
\includegraphics[width=1\linewidth]{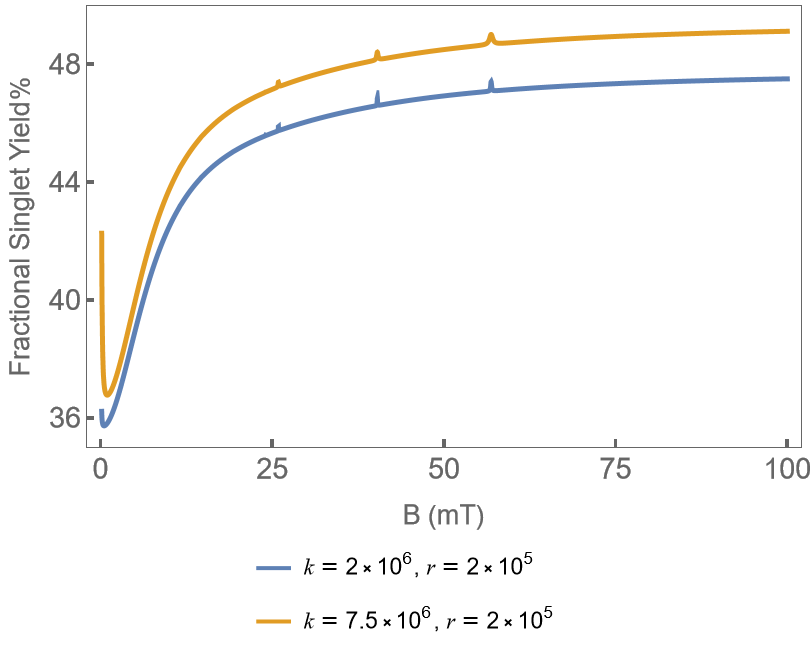}   
  \caption{}
  \label{fig:1.3}
\end{subfigure}
\caption{(a) Ratio of the FSY percentages, $S$ plotted in the $k-r$ plane for the Ser Oxyradical case. The lighter yellow areas are closest in value to the ratios seen in Tables \ref{tab1} \& \ref{tab2}. (b) Variation of the FSY percentage with magnetic field strength ranging from $0.15$ mT to $100$ mT. The figure is plotted keeping $r = 2\times10^5\mbox{ s}^{-1}$. The reaction rates are $k =  2\times 10^6\mbox{ s}^{-1}$ and $k =  7.5\times 10^6\mbox{ s}^{-1}$, which correspond to $S = 1.32$ and $S = 1.17$. All the plots are plotted considering the HFI contribution from the hydrogen atom of the Serine oxyradical with the highest HFCC, $a_A = 7.45$ mT. We also consider the contribution of the $^{25}$Mg isotope at its natural abundance of 10\%. The HFI contribution from the same is $a_B = -11.22$ mT.}
\label{fig:1}
\end{figure}

Fig.\ref{fig:1.1} depicts a plot of $S$ versus $k$ and $r$ for the radical pair containing the serine oxyradical. In this calculation, we consider the contribution of all Magnesium isotopes according to their natural abundance. As the magnetic field strength for the control setup of the experiment conducted by Hirai et al.\cite{Hirai2003, Hirai2005} ranged from $B_0 = 0-0.3$ mT, we take $B_0 = 0.15$ mT as our control value. Here, the lighter yellow colored areas of the plot within the black outline represents the values of $k$ and $r$ for which there is the greatest variation in the FSY between $B_0$ and $B_{\mbox{exp}}$. These points have the potential to capture the behaviour of the system, as they match the magnitude of the ratio of $\mbox{Ca}^{2+}$ concentration in the control vs. experimental setups. The values of $k$ and $r$ are consistent with typical values considered in the literature on the RPM.

Examples of the variation of FSY percentage with magnetic field strength from $0.15\mbox{ mT}$ to $100\mbox{ mT}$ for two different choice of $k$ and $r$ from the feasible range of values is given in Figs. \ref{fig:1}.  

\subsubsection*{Tyr Pathway}
\begin{figure}[h!]
\begin{subfigure}{.49\textwidth}
  \centering
  % include first image
\includegraphics[width=1\linewidth]{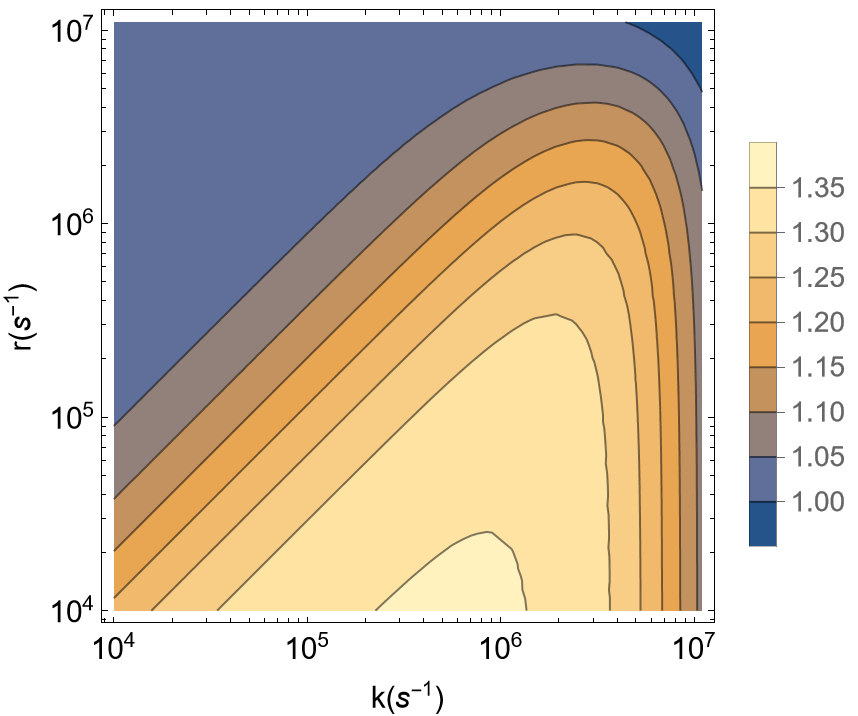}
  \caption{}
  \label{fig:2.1}
\end{subfigure}
\begin{subfigure}{.49\textwidth}
  \centering
  % include second image
\includegraphics[width=1\linewidth]{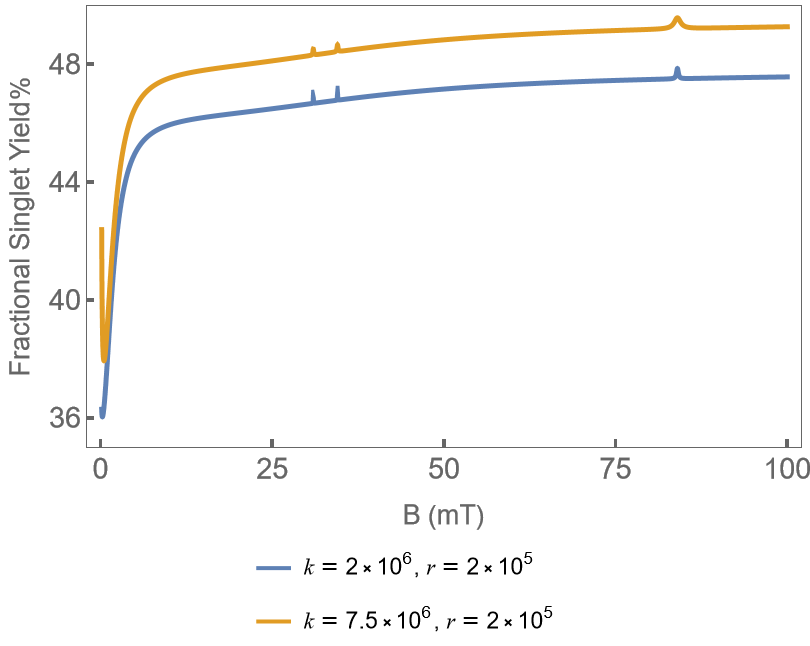}   
  % \caption{}
  \label{fig:2.2}
\end{subfigure}
\caption{(a) Ratio of the FSY percentages, $S$ plotted in the $k-r$ plane. The lighter yellow areas are closer in value to the ratios seen in Tables \ref{tab1} \& \ref{tab2}. (b) Variation of the FSY percentage with magnetic field strength ranging from $0.15$ mT to $100$ mT. The figure is plotted keeping $r = 2\times10^5\mbox{ s}^{-1}$. The reaction rates are $k =  2\times 10^6\mbox{ s}^{-1}$ and $k =  7.5\times 10^6\mbox{ s}^{-1}$, which correspond to $S = 1.32$ and $S = 1.17$. All the figures are plotted considering the HFI contribution from the hydrogen atom of the Tyrosine oxyradical with the highest HFCC, $a_A = 1.86$ mT. We also consider the contribution of the $^{25}$Mg isotope at its natural abundance of 10\%. The HFI contribution from the same is $a_B = -11.22$ mT.}
\label{fig:2}
\end{figure}

Similar to what we have done for the case of the Ser Pathway, the quantity $S$ was calculated and plotted against phyically feasible ranges of $k$ and $r$ for the radical pair involving the tyrosine oxyradical. Fig. \ref{fig:2.1} shows the same. 

Fig. \ref{fig:2} also shows examples of the variation of FSY percentage with magnetic field strength from $0.15\mbox{ mT}$ to $100\mbox{ mT}$ for two different values of $k$ and $r$.

\hfill\break
\noindent
The FSY was found to exhibit a few spikes at certain values of the MF strength, B, for both the Ser and Tyr Pathway. These spikes are associated with certain eigenstates overlapping in energy due to the effect of the spin of the $^{25}$Mg isotope. A dip in the FSY values near zero was also found at higher values of $k$, as seen from the examples given in Figs. \ref{fig:1} \& \ref{fig:2}. The extent of the drop increases as $k$ increases for both Ser and Tyr Pathways. The drop in value is seen in the behaviour of the system under the effect of both isotopes. More results regarding the contribution from the individual isotopes are presented in the Appendix of the paper.

\section*{Discussion}

The primary objective of this study was to determine whether the radical pair mechanism can provide a plausible explanation for the magnetic field effects on NMDAR activity reported by Hirai et al.\cite{Hirai2003, Hirai2005}. We find that a simple RP model is capable of capturing the MFEs on the system, as the magnitude of the MFE on the singlet yield corresponds well to the magnitude of the experimental MFE observed on the $\mbox{Ca}^{2+}$ ion concentration within the cells. It has been shown above that both Serine and Tyrosine phosphorylation could be involved in the upregulation of NMDAR activity in the presence of a static magnetic field. Serine phosphorylation is involved in activating the AP1 transcription factor, which is a major regulator of the NR2B subunit. As described previously, an increase in NR2B can increase the number of NMDARs in the system, resulting in an increase in the calcium ion influx into the cells. We have also seen how phosphorylation of certain NMDAR Subunits at Serine and Tyrosine sites may also cause an increase in NMDAR activity, which again raises the possibility that magnetic field effects occur through the phosphorylation processes. We thus propose that the radical pairs involved in the ion-radical mechanism of phosphorylation could be a likely candidate for the RPM model. We thus consider radical pairs of the form $[\mbox{RO}^\bullet \mbox{ Mg($\mbox{H}_2$O$)_n$}^{+\bullet}]$, where the protein residue, R could be a Tyrosine or Serine residue.

Based on the experimental evidence found by Hirai et al., there is an increase in the concentration of $\mbox{Ca}^{2+}$ ions within the hippocampal cells in the presence of magnetic fields. For cells exposed to a sustained or repetitive SMF, an increase in concentration of the ions by an average factor of about 1.5 and 1.4, respectively, has been observed. So far, our results indicate that this variation in the behavior of the system in the presence of the magnetic field could indeed be explained using an RPM model. 

The radical pairs involving both the Serine and Tyrosine pathway models exhibit significant variation in singlet yield percentage in the presence of the magnetic field. We find that both models capture a variation in FSY by a factor of over 1.3 in the presence of the 100 mT magnetic field for certain plausible values of $k$ and $r$. Based on these results, we propose that the increase in the yield of the singlet product of the reaction could be leading to the increase in the $\mbox{Ca}^{2+}$ ion concentration through various biochemical reactions within the system. Though the increase of the singlet yield of the model is not exactly equal to the experimentally observed increase in the $\mbox{Ca}^{2+}$ concentration, it is plausible that an amplification of the effects could occur through the various processes involved in the $\mbox{Ca}^{2+}$ ion intake into the cells. Multiple studies have shown that such amplifications are a promising possibility in numerous systems involving the RPM model\cite{Kattnig2016, Kattnig2017, Player2021, Binhi2023}. Therefore, we propose that both the Ser and Tyr Pathways could potentially explain the MFE observed in experiments. However, there is limited evidence of the formation of serine radicals, particularly in biological systems\cite{Forbes2007,Heo2006,Liming1969}. It is worth noting that the radical pairs under consideration could also be involved in various other reactions within the biological system, which could also play a role in the observed magnetic field effects.

Based on the results we have presented, we propose that the radical pair mechanism is a promising model to explain the effects of the magnetic field effects on NMDAR activity. New experiments studying the effect of lower magnetic field strength on NMDA activity could give further insight into the behaviour of the system and the accuracy of the results from our model. Further experiments could also be conducted to study the likelihood of Serine and Tyrosine radical formation in the system and the role they could play in the pathways leading to the MFE on NMDA activity. The most common method for radical detection in biological systems is EPR spectroscopy. Direct EPR spectroscopy could be used to detect the presence of the Tyrosine radical as they are long-lived due to stabilization by electron delocalization. There is some evidence of EPR and time-resolved EPR being successfully used to study Serine radicals\cite{Forbes2007,Liming1969}. These could potentially be used to study if these radicals are present in a given system. Experiments that study the effects of the magnetic field in a system with a higher concentration of the $^{25}$Mg isotope could also be interesting, as we find certain isotope dependent effects in the calculation of FSY, which is discussed in detail in the Appendix section.

It is possible that similar models based on the RPM could be used to understand magnetic field effects on various other biological processes. Furthering our understanding of how external stimuli like magnetic fields affect the different processes within the brain could lead to a much deeper understanding of its working. It could also provide significant insights into finding ways to diagnose and treat various neuropsychiatric and psychological disorders, as well, thus making this simultaneously an intriguing and practically relevant field of study.

\section*{Methods}

\subsection*{DFT Analysis}

The DFT calculations for our radical pairs were performed using the ORCA package\cite{Neese2012}. The optimization of the molecular structures was performed using the dispersion-corrected PBE0 functional and the def2-TZVP basis set.

The hyperfine coupling constants were calculated from the orbitals obtained from the optimization calculations. The B3LYP functional and def2-TZVP basis set were used for the calculation of both the HFCCs, $a_A$, and $a_B$. The solvent effects in every calculation were considered using the conductor-like polarizable continuum model (CPCM) with a dielectric constant of 2.

\bibliography{Refs}

\begin{thebibliography}{10}
\urlstyle{rm}
\expandafter\ifx\csname url\endcsname\relax
  \def\url#1{\texttt{#1}}\fi
\expandafter\ifx\csname urlprefix\endcsname\relax\def\urlprefix{URL }\fi
\expandafter\ifx\csname doiprefix\endcsname\relax\def\doiprefix{DOI: }\fi
\providecommand{\bibinfo}[2]{#2}
\providecommand{\eprint}[2][]{\url{#2}}

\bibitem{Johnsen2005}
\bibinfo{author}{Johnsen, S.} \& \bibinfo{author}{Lohmann, K.~J.}
\newblock \bibinfo{journal}{\bibinfo{title}{The physics and neurobiology of
  magnetoreception}}.
\newblock {\emph{\JournalTitle{Nature Reviews Neuroscience}}}
  \textbf{\bibinfo{volume}{6}}, \bibinfo{pages}{703--712},
  \doiprefix\url{10.1038/nrn1745} (\bibinfo{year}{2005}).

\bibitem{Lewczuk2014}
\bibinfo{author}{Lewczuk, B.} \emph{et~al.}
\newblock \bibinfo{journal}{\bibinfo{title}{Influence of electric, magnetic,
  and electromagnetic fields on the circadian system: Current stage of
  knowledge}}.
\newblock {\emph{\JournalTitle{{BioMed} Research International}}}
  \textbf{\bibinfo{volume}{2014}}, \bibinfo{pages}{1--13},
  \doiprefix\url{10.1155/2014/169459} (\bibinfo{year}{2014}).

\bibitem{McCann1998}
\bibinfo{author}{McCann, J.}, \bibinfo{author}{Dietrich, F.} \&
  \bibinfo{author}{Rafferty, C.}
\newblock \bibinfo{journal}{\bibinfo{title}{The genotoxic potential of electric
  and magnetic fields: an update}}.
\newblock {\emph{\JournalTitle{Mutation Research/Reviews in Mutation
  Research}}} \textbf{\bibinfo{volume}{411}}, \bibinfo{pages}{45--86},
  \doiprefix\url{10.1016/s1383-5742(98)00006-4} (\bibinfo{year}{1998}).

\bibitem{Ketchen1978}
\bibinfo{author}{KETCHEN, E.}, \bibinfo{author}{PORTER, W.} \&
  \bibinfo{author}{BOLTON, N.}
\newblock \bibinfo{journal}{\bibinfo{title}{The biological effects of magnetic
  fields on man}}.
\newblock {\emph{\JournalTitle{American Industrial Hygiene Association
  Journal}}} \textbf{\bibinfo{volume}{39}}, \bibinfo{pages}{1--11},
  \doiprefix\url{10.1080/0002889778507706} (\bibinfo{year}{1978}).

\bibitem{Fan2021}
\bibinfo{author}{Fan, Y.}, \bibinfo{author}{Ji, X.}, \bibinfo{author}{Zhang,
  L.} \& \bibinfo{author}{Zhang, X.}
\newblock \bibinfo{journal}{\bibinfo{title}{The analgesic effects of static
  magnetic fields}}.
\newblock {\emph{\JournalTitle{Bioelectromagnetics}}}
  \textbf{\bibinfo{volume}{42}}, \bibinfo{pages}{115--127},
  \doiprefix\url{10.1002/bem.22323} (\bibinfo{year}{2021}).

\bibitem{ZadehHaghighi2022}
\bibinfo{author}{Zadeh-Haghighi, H.} \& \bibinfo{author}{Simon, C.}
\newblock \bibinfo{journal}{\bibinfo{title}{Magnetic field effects in biology
  from the perspective of the radical pair mechanism}}.
\newblock {\emph{\JournalTitle{Journal of The Royal Society Interface}}}
  \textbf{\bibinfo{volume}{19}}, \doiprefix\url{10.1098/rsif.2022.0325}
  (\bibinfo{year}{2022}).

\bibitem{Buchachenko2014}
\bibinfo{author}{Buchachenko, A.~L.}
\newblock \bibinfo{journal}{\bibinfo{title}{Magnetic field-dependent molecular
  and chemical processes in biochemistry, genetics and medicine}}.
\newblock {\emph{\JournalTitle{Russian Chemical Reviews}}}
  \textbf{\bibinfo{volume}{83}}, \bibinfo{pages}{1--12},
  \doiprefix\url{10.1070/rc2014v083n01abeh004335} (\bibinfo{year}{2014}).

\bibitem{Albuquerque2016}
\bibinfo{author}{Albuquerque, W. W.~C.}, \bibinfo{author}{Costa, R. M. P.~B.},
  \bibinfo{author}{de~Salazar~e Fernandes, T.} \& \bibinfo{author}{Porto, A.
  L.~F.}
\newblock \bibinfo{journal}{\bibinfo{title}{Evidences of the static magnetic
  field influence on cellular systems}}.
\newblock {\emph{\JournalTitle{Progress in Biophysics and Molecular Biology}}}
  \textbf{\bibinfo{volume}{121}}, \bibinfo{pages}{16--28},
  \doiprefix\url{10.1016/j.pbiomolbio.2016.03.003} (\bibinfo{year}{2016}).

\bibitem{McKay2007}
\bibinfo{author}{McKay, J.~C.}, \bibinfo{author}{Prato, F.~S.} \&
  \bibinfo{author}{Thomas, A.~W.}
\newblock \bibinfo{journal}{\bibinfo{title}{A literature review: The effects of
  magnetic field exposure on blood flow and blood vessels in the
  microvasculature}}.
\newblock {\emph{\JournalTitle{Bioelectromagnetics}}}
  \textbf{\bibinfo{volume}{28}}, \bibinfo{pages}{81--98},
  \doiprefix\url{10.1002/bem.20284} (\bibinfo{year}{2007}).

\bibitem{Yang2023}
\bibinfo{author}{Yang, J.}, \bibinfo{author}{Feng, Y.}, \bibinfo{author}{Li,
  Q.} \& \bibinfo{author}{Zeng, Y.}
\newblock \bibinfo{journal}{\bibinfo{title}{Evidence of the static magnetic
  field effects on bone-related diseases and bone cells}}.
\newblock {\emph{\JournalTitle{Progress in Biophysics and Molecular Biology}}}
  \textbf{\bibinfo{volume}{177}}, \bibinfo{pages}{168--180},
  \doiprefix\url{10.1016/j.pbiomolbio.2022.11.006} (\bibinfo{year}{2023}).

\bibitem{Adair2000}
\bibinfo{author}{Adair, R.~K.}
\newblock \bibinfo{journal}{\bibinfo{title}{Static and low-frequency magnetic
  field effects: health risks and therapies}}.
\newblock {\emph{\JournalTitle{Reports on Progress in Physics}}}
  \textbf{\bibinfo{volume}{63}}, \bibinfo{pages}{415--454},
  \doiprefix\url{10.1088/0034-4885/63/3/204} (\bibinfo{year}{2000}).

\bibitem{Bingi2003}
\bibinfo{author}{Bingi, V.~N.} \& \bibinfo{author}{Savin, A.~V.}
\newblock \bibinfo{journal}{\bibinfo{title}{Effects of weak magnetic fields on
  biological systems: physical aspects}}.
\newblock {\emph{\JournalTitle{Physics-Uspekhi}}}
  \textbf{\bibinfo{volume}{46}}, \bibinfo{pages}{259--291},
  \doiprefix\url{10.1070/pu2003v046n03abeh001283} (\bibinfo{year}{2003}).

\bibitem{Maffei2014}
\bibinfo{author}{Maffei, M.~E.}
\newblock \bibinfo{journal}{\bibinfo{title}{Magnetic field effects on plant
  growth, development, and evolution}}.
\newblock {\emph{\JournalTitle{Frontiers in Plant Science}}}
  \textbf{\bibinfo{volume}{5}}, \doiprefix\url{10.3389/fpls.2014.00445}
  (\bibinfo{year}{2014}).

\bibitem{Karimi2020}
\bibinfo{author}{Karimi, A.}, \bibinfo{author}{Moghaddam, F.~G.} \&
  \bibinfo{author}{Valipour, M.}
\newblock \bibinfo{journal}{\bibinfo{title}{Insights in the biology of
  extremely low-frequency magnetic fields exposure on human health}}.
\newblock {\emph{\JournalTitle{Molecular Biology Reports}}}
  \textbf{\bibinfo{volume}{47}}, \bibinfo{pages}{5621--5633},
  \doiprefix\url{10.1007/s11033-020-05563-8} (\bibinfo{year}{2020}).

\bibitem{2018}
\bibinfo{editor}{Barnes, F.~S.} \& \bibinfo{editor}{Greenebaum, B.} (eds.)
  \emph{\bibinfo{title}{Handbook of Biological Effects of Electromagnetic
  Fields - Two Volume Set}} (\bibinfo{publisher}{{CRC} Press},
  \bibinfo{year}{2018}).

\bibitem{Kim2021}
\bibinfo{author}{Kim, Y.} \emph{et~al.}
\newblock \bibinfo{journal}{\bibinfo{title}{Quantum biology: An update and
  perspective}}.
\newblock {\emph{\JournalTitle{Quantum Reports}}} \textbf{\bibinfo{volume}{3}},
  \bibinfo{pages}{80--126}, \doiprefix\url{10.3390/quantum3010006}
  (\bibinfo{year}{2021}).

\bibitem{Timmel1998}
\bibinfo{author}{Timmel, C.}, \bibinfo{author}{Till, U.},
  \bibinfo{author}{Brocklehurst, B.}, \bibinfo{author}{Mclauchlan, K.} \&
  \bibinfo{author}{Hore, P.}
\newblock \bibinfo{journal}{\bibinfo{title}{Effects of weak magnetic fields on
  free radical recombination reactions}}.
\newblock {\emph{\JournalTitle{Molecular Physics}}}
  \textbf{\bibinfo{volume}{95}}, \bibinfo{pages}{71--89},
  \doiprefix\url{10.1080/00268979809483134} (\bibinfo{year}{1998}).

\bibitem{Hore2016}
\bibinfo{author}{Hore, P.~J.} \& \bibinfo{author}{Mouritsen, H.}
\newblock \bibinfo{journal}{\bibinfo{title}{The radical-pair mechanism of
  magnetoreception}}.
\newblock {\emph{\JournalTitle{Annual Review of Biophysics}}}
  \textbf{\bibinfo{volume}{45}}, \bibinfo{pages}{299--344},
  \doiprefix\url{10.1146/annurev-biophys-032116-094545} (\bibinfo{year}{2016}).

\bibitem{Mouritsen2018}
\bibinfo{author}{Mouritsen, H.}
\newblock \bibinfo{journal}{\bibinfo{title}{Long-distance navigation and
  magnetoreception in migratory animals}}.
\newblock {\emph{\JournalTitle{Nature}}} \textbf{\bibinfo{volume}{558}},
  \bibinfo{pages}{50--59}, \doiprefix\url{10.1038/s41586-018-0176-1}
  (\bibinfo{year}{2018}).

\bibitem{Simons2005}
\bibinfo{author}{Simons, W.} \& \bibinfo{author}{Dierick, M.}
\newblock \bibinfo{journal}{\bibinfo{title}{Transcranial magnetic stimulation
  as a therapeutic tool in psychiatry}}.
\newblock {\emph{\JournalTitle{The World Journal of Biological Psychiatry}}}
  \textbf{\bibinfo{volume}{6}}, \bibinfo{pages}{6--25},
  \doiprefix\url{10.1080/15622970510029812} (\bibinfo{year}{2005}).

\bibitem{Antczak2021}
\bibinfo{author}{Antczak, J.}, \bibinfo{author}{Rusin, G.} \&
  \bibinfo{author}{S{\l}owik, A.}
\newblock \bibinfo{journal}{\bibinfo{title}{Transcranial magnetic stimulation
  as a diagnostic and therapeutic tool in various types of dementia}}.
\newblock {\emph{\JournalTitle{Journal of Clinical Medicine}}}
  \textbf{\bibinfo{volume}{10}}, \bibinfo{pages}{2875},
  \doiprefix\url{10.3390/jcm10132875} (\bibinfo{year}{2021}).

\bibitem{MAntczak2021}
\bibinfo{author}{Antczak, J.~M.}
\newblock \bibinfo{journal}{\bibinfo{title}{Transcranial magnetic stimulation
  as a~diagnostic and therapeutic tool in cerebral palsy}}.
\newblock {\emph{\JournalTitle{Advances in Psychiatry and Neurology}}}
  \textbf{\bibinfo{volume}{30}}, \bibinfo{pages}{203--212},
  \doiprefix\url{10.5114/ppn.2021.110796} (\bibinfo{year}{2021}).

\bibitem{Luber2013}
\bibinfo{author}{Luber, B.}, \bibinfo{author}{McClintock, S.~M.} \&
  \bibinfo{author}{Lisanby, S.~H.}
\newblock \bibinfo{journal}{\bibinfo{title}{Applications of transcranial
  magnetic stimulation and magnetic seizure therapy in the study and treatment
  of disorders related to cerebral aging}}.
\newblock {\emph{\JournalTitle{Dialogues in Clinical Neuroscience}}}
  \textbf{\bibinfo{volume}{15}}, \bibinfo{pages}{87--98},
  \doiprefix\url{10.31887/dcns.2013.15.1/bluber} (\bibinfo{year}{2013}).

\bibitem{NOJIMA2020}
\bibinfo{author}{NOJIMA, I.}, \bibinfo{author}{OLIVIERO, A.} \&
  \bibinfo{author}{MIMA, T.}
\newblock \bibinfo{journal}{\bibinfo{title}{Transcranial static magnetic
  stimulation {\textemdash}from bench to bedside and beyond{\textemdash}}}.
\newblock {\emph{\JournalTitle{Neuroscience Research}}}
  \textbf{\bibinfo{volume}{156}}, \bibinfo{pages}{250--255},
  \doiprefix\url{10.1016/j.neures.2019.12.005} (\bibinfo{year}{2020}).

\bibitem{ZadehHaghighi2021}
\bibinfo{author}{Zadeh-Haghighi, H.} \& \bibinfo{author}{Simon, C.}
\newblock \bibinfo{journal}{\bibinfo{title}{Entangled radicals may explain
  lithium effects on hyperactivity}}.
\newblock {\emph{\JournalTitle{Scientific Reports}}}
  \textbf{\bibinfo{volume}{11}}, \doiprefix\url{10.1038/s41598-021-91388-9}
  (\bibinfo{year}{2021}).

\bibitem{ZHaghighi2022}
\bibinfo{author}{Zadeh-Haghighi, H.} \& \bibinfo{author}{Simon, C.}
\newblock \bibinfo{journal}{\bibinfo{title}{Radical pairs can explain magnetic
  field and lithium effects on the circadian clock}}.
\newblock {\emph{\JournalTitle{Scientific Reports}}}
  \textbf{\bibinfo{volume}{12}}, \doiprefix\url{10.1038/s41598-021-04334-0}
  (\bibinfo{year}{2022}).

\bibitem{Rishabh2022}
\bibinfo{author}{Rishabh, R.}, \bibinfo{author}{Zadeh-Haghighi, H.},
  \bibinfo{author}{Salahub, D.} \& \bibinfo{author}{Simon, C.}
\newblock \bibinfo{journal}{\bibinfo{title}{Radical pairs may explain reactive
  oxygen species-mediated effects of hypomagnetic field on neurogenesis}}.
\newblock {\emph{\JournalTitle{{PLOS} Computational Biology}}}
  \textbf{\bibinfo{volume}{18}}, \bibinfo{pages}{e1010198},
  \doiprefix\url{10.1371/journal.pcbi.1010198} (\bibinfo{year}{2022}).

\bibitem{ZH2022}
\bibinfo{author}{Zadeh-Haghighi, H.} \& \bibinfo{author}{Simon, C.}
\newblock \bibinfo{journal}{\bibinfo{title}{Radical pairs may play a role in
  microtubule reorganization}}.
\newblock {\emph{\JournalTitle{Scientific Reports}}}
  \textbf{\bibinfo{volume}{12}}, \doiprefix\url{10.1038/s41598-022-10068-4}
  (\bibinfo{year}{2022}).

\bibitem{Smith2021}
\bibinfo{author}{Smith, J.}, \bibinfo{author}{Haghighi, H.~Z.},
  \bibinfo{author}{Salahub, D.} \& \bibinfo{author}{Simon, C.}
\newblock \bibinfo{journal}{\bibinfo{title}{Radical pairs may play a role in
  xenon-induced general anesthesia}}.
\newblock {\emph{\JournalTitle{Scientific Reports}}}
  \textbf{\bibinfo{volume}{11}}, \doiprefix\url{10.1038/s41598-021-85673-w}
  (\bibinfo{year}{2021}).

\bibitem{zgn2019}
\bibinfo{author}{\"{O}zg\"{u}n, A.}, \bibinfo{author}{Marote, A.},
  \bibinfo{author}{Behie, L.~A.}, \bibinfo{author}{Salgado, A.} \&
  \bibinfo{author}{Garipcan, B.}
\newblock \bibinfo{journal}{\bibinfo{title}{Extremely low frequency magnetic
  field induces human neuronal differentiation through {NMDA} receptor
  activation}}.
\newblock {\emph{\JournalTitle{Journal of Neural Transmission}}}
  \textbf{\bibinfo{volume}{126}}, \bibinfo{pages}{1281--1290},
  \doiprefix\url{10.1007/s00702-019-02045-5} (\bibinfo{year}{2019}).

\bibitem{Salunke2013}
\bibinfo{author}{Salunke, B.~P.}, \bibinfo{author}{Umathe, S.~N.} \&
  \bibinfo{author}{Chavan, J.~G.}
\newblock \bibinfo{journal}{\bibinfo{title}{Involvement of {NMDA} receptor in
  low-frequency magnetic field-induced anxiety in mice}}.
\newblock {\emph{\JournalTitle{Electromagnetic Biology and Medicine}}}
  \textbf{\bibinfo{volume}{33}}, \bibinfo{pages}{312--326},
  \doiprefix\url{10.3109/15368378.2013.839453} (\bibinfo{year}{2013}).

\bibitem{Paoletti2013}
\bibinfo{author}{Paoletti, P.}, \bibinfo{author}{Bellone, C.} \&
  \bibinfo{author}{Zhou, Q.}
\newblock \bibinfo{journal}{\bibinfo{title}{{NMDA} receptor subunit diversity:
  impact on receptor properties, synaptic plasticity and disease}}.
\newblock {\emph{\JournalTitle{Nature Reviews Neuroscience}}}
  \textbf{\bibinfo{volume}{14}}, \bibinfo{pages}{383--400},
  \doiprefix\url{10.1038/nrn3504} (\bibinfo{year}{2013}).

\bibitem{SanzClemente2012}
\bibinfo{author}{Sanz-Clemente, A.}, \bibinfo{author}{Nicoll, R.~A.} \&
  \bibinfo{author}{Roche, K.~W.}
\newblock \bibinfo{journal}{\bibinfo{title}{Diversity in {NMDA} receptor
  composition}}.
\newblock {\emph{\JournalTitle{The Neuroscientist}}}
  \textbf{\bibinfo{volume}{19}}, \bibinfo{pages}{62--75},
  \doiprefix\url{10.1177/1073858411435129} (\bibinfo{year}{2012}).

\bibitem{PAOLETTI2007}
\bibinfo{author}{PAOLETTI, P.} \& \bibinfo{author}{NEYTON, J.}
\newblock \bibinfo{journal}{\bibinfo{title}{{NMDA} receptor subunits: function
  and pharmacology}}.
\newblock {\emph{\JournalTitle{Current Opinion in Pharmacology}}}
  \textbf{\bibinfo{volume}{7}}, \bibinfo{pages}{39--47},
  \doiprefix\url{10.1016/j.coph.2006.08.011} (\bibinfo{year}{2007}).

\bibitem{Monyer1992}
\bibinfo{author}{Monyer, H.} \emph{et~al.}
\newblock \bibinfo{journal}{\bibinfo{title}{Heteromeric {NMDA} receptors:
  Molecular and functional distinction of subtypes}}.
\newblock {\emph{\JournalTitle{Science}}} \textbf{\bibinfo{volume}{256}},
  \bibinfo{pages}{1217--1221}, \doiprefix\url{10.1126/science.256.5060.1217}
  (\bibinfo{year}{1992}).

\bibitem{Furukawa2005}
\bibinfo{author}{Furukawa, H.}, \bibinfo{author}{Singh, S.~K.},
  \bibinfo{author}{Mancusso, R.} \& \bibinfo{author}{Gouaux, E.}
\newblock \bibinfo{journal}{\bibinfo{title}{Subunit arrangement and function in
  {NMDA} receptors}}.
\newblock {\emph{\JournalTitle{Nature}}} \textbf{\bibinfo{volume}{438}},
  \bibinfo{pages}{185--192}, \doiprefix\url{10.1038/nature04089}
  (\bibinfo{year}{2005}).

\bibitem{Hansen2017}
\bibinfo{author}{Hansen, K.~B.}, \bibinfo{author}{Yi, F.},
  \bibinfo{author}{Perszyk, R.~E.}, \bibinfo{author}{Menniti, F.~S.} \&
  \bibinfo{author}{Traynelis, S.~F.}
\newblock \bibinfo{title}{{NMDA} receptors in the central nervous system}.
\newblock In \emph{\bibinfo{booktitle}{Methods in Molecular Biology}},
  \bibinfo{pages}{1--80}, \doiprefix\url{10.1007/978-1-4939-7321-7_1}
  (\bibinfo{publisher}{Springer New York}, \bibinfo{year}{2017}).

\bibitem{Liu2019}
\bibinfo{author}{Liu, J.}, \bibinfo{author}{Chang, L.}, \bibinfo{author}{Song,
  Y.}, \bibinfo{author}{Li, H.} \& \bibinfo{author}{Wu, Y.}
\newblock \bibinfo{journal}{\bibinfo{title}{The role of {NMDA} receptors in
  alzheimer's disease}}.
\newblock {\emph{\JournalTitle{Frontiers in Neuroscience}}}
  \textbf{\bibinfo{volume}{13}}, \doiprefix\url{10.3389/fnins.2019.00043}
  (\bibinfo{year}{2019}).

\bibitem{Nakazawa2020}
\bibinfo{author}{Nakazawa, K.} \& \bibinfo{author}{Sapkota, K.}
\newblock \bibinfo{journal}{\bibinfo{title}{The origin of {NMDA} receptor
  hypofunction in schizophrenia}}.
\newblock {\emph{\JournalTitle{Pharmacology \& Therapeutics}}}
  \textbf{\bibinfo{volume}{205}}, \bibinfo{pages}{107426},
  \doiprefix\url{10.1016/j.pharmthera.2019.107426} (\bibinfo{year}{2020}).

\bibitem{Zhou2013}
\bibinfo{author}{Zhou, Q.} \& \bibinfo{author}{Sheng, M.}
\newblock \bibinfo{journal}{\bibinfo{title}{{NMDA} receptors in nervous system
  diseases}}.
\newblock {\emph{\JournalTitle{Neuropharmacology}}}
  \textbf{\bibinfo{volume}{74}}, \bibinfo{pages}{69--75},
  \doiprefix\url{10.1016/j.neuropharm.2013.03.030} (\bibinfo{year}{2013}).

\bibitem{Hirai2003}
\bibinfo{author}{Hirai, T.} \& \bibinfo{author}{Yoneda, Y.}
\newblock \bibinfo{journal}{\bibinfo{title}{Functional alterations in immature
  cultured rat hippocampal neurons after sustained exposure to static magnetic
  fields}}.
\newblock {\emph{\JournalTitle{Journal of Neuroscience Research}}}
  \textbf{\bibinfo{volume}{75}}, \bibinfo{pages}{230--240},
  \doiprefix\url{10.1002/jnr.10819} (\bibinfo{year}{2003}).

\bibitem{Hirai2005}
\bibinfo{author}{Hirai, T.} \emph{et~al.}
\newblock \bibinfo{journal}{\bibinfo{title}{Counteraction by repetitive daily
  exposure to static magnetism against sustained blockade of
  n-methyl-d-aspartate receptor channels in cultured rat hippocampal neurons}}.
\newblock {\emph{\JournalTitle{Journal of Neuroscience Research}}}
  \textbf{\bibinfo{volume}{80}}, \bibinfo{pages}{491--500},
  \doiprefix\url{10.1002/jnr.20497} (\bibinfo{year}{2005}).

\bibitem{Qiang2005}
\bibinfo{author}{Qiang, M.} \& \bibinfo{author}{Ticku, M.~K.}
\newblock \bibinfo{journal}{\bibinfo{title}{Role of {AP}-1 in ethanol-induced
  n-methyl-d-aspartate receptor 2b subunit gene up-regulation in mouse cortical
  neurons}}.
\newblock {\emph{\JournalTitle{Journal of Neurochemistry}}}
  \textbf{\bibinfo{volume}{95}}, \bibinfo{pages}{1332--1341},
  \doiprefix\url{10.1111/j.1471-4159.2005.03464.x} (\bibinfo{year}{2005}).

\bibitem{Prybylowski2002}
\bibinfo{author}{Prybylowski, K.} \emph{et~al.}
\newblock \bibinfo{journal}{\bibinfo{title}{Relationship between availability
  of {NMDA} receptor subunits and their expression at the synapse}}.
\newblock {\emph{\JournalTitle{The Journal of Neuroscience}}}
  \textbf{\bibinfo{volume}{22}}, \bibinfo{pages}{8902--8910},
  \doiprefix\url{10.1523/jneurosci.22-20-08902.2002} (\bibinfo{year}{2002}).

\bibitem{Gupta1996}
\bibinfo{author}{Gupta, S.} \emph{et~al.}
\newblock \bibinfo{journal}{\bibinfo{title}{Selective interaction of {JNK}
  protein kinase isoforms with transcription factors.}}
\newblock {\emph{\JournalTitle{The {EMBO} Journal}}}
  \textbf{\bibinfo{volume}{15}}, \bibinfo{pages}{2760--2770},
  \doiprefix\url{10.1002/j.1460-2075.1996.tb00636.x} (\bibinfo{year}{1996}).

\bibitem{Kallunki1996}
\bibinfo{author}{Kallunki, T.}, \bibinfo{author}{Deng, T.},
  \bibinfo{author}{Hibi, M.} \& \bibinfo{author}{Karin, M.}
\newblock \bibinfo{journal}{\bibinfo{title}{c-jun can recruit {JNK} to
  phosphorylate dimerization partners via specific docking interactions}}.
\newblock {\emph{\JournalTitle{Cell}}} \textbf{\bibinfo{volume}{87}},
  \bibinfo{pages}{929--939}, \doiprefix\url{10.1016/s0092-8674(00)81999-6}
  (\bibinfo{year}{1996}).

\bibitem{Markov1997}
\bibinfo{author}{Markov, M.} \& \bibinfo{author}{Pilla, A.}
\newblock \bibinfo{journal}{\bibinfo{title}{Weak static magnetic field
  modulation of myosin phosphorylation in a cell-free preparation: Calcium
  dependence}}.
\newblock {\emph{\JournalTitle{Bioelectrochemistry and Bioenergetics}}}
  \textbf{\bibinfo{volume}{43}}, \bibinfo{pages}{233--238},
  \doiprefix\url{10.1016/s0302-4598(96)02226-x} (\bibinfo{year}{1997}).

\bibitem{Engstrm2002}
\bibinfo{author}{Engstr\"{o}m, S.}, \bibinfo{author}{Markov, M.~S.},
  \bibinfo{author}{McLean, M.~J.}, \bibinfo{author}{Holcomb, R.~R.} \&
  \bibinfo{author}{Markov, J.~M.}
\newblock \bibinfo{journal}{\bibinfo{title}{Effects of non-uniform static
  magnetic fields on the rate of myosin phosphorylation}}.
\newblock {\emph{\JournalTitle{Bioelectromagnetics}}}
  \textbf{\bibinfo{volume}{23}}, \bibinfo{pages}{475--479},
  \doiprefix\url{10.1002/bem.10035} (\bibinfo{year}{2002}).

\bibitem{Buchachenko2005}
\bibinfo{author}{Buchachenko, A.~L.}, \bibinfo{author}{Kouznetsov, D.~A.},
  \bibinfo{author}{Orlova, M.~A.} \& \bibinfo{author}{Markarian, A.~A.}
\newblock \bibinfo{journal}{\bibinfo{title}{Magnetic isotope effect of
  magnesium in phosphoglycerate kinase phosphorylation}}.
\newblock {\emph{\JournalTitle{Proceedings of the National Academy of
  Sciences}}} \textbf{\bibinfo{volume}{102}}, \bibinfo{pages}{10793--10796},
  \doiprefix\url{10.1073/pnas.0504876102} (\bibinfo{year}{2005}).

\bibitem{Chen2007}
\bibinfo{author}{Chen, B.-S.} \& \bibinfo{author}{Roche, K.~W.}
\newblock \bibinfo{journal}{\bibinfo{title}{Regulation of {NMDA} receptors by
  phosphorylation}}.
\newblock {\emph{\JournalTitle{Neuropharmacology}}}
  \textbf{\bibinfo{volume}{53}}, \bibinfo{pages}{362--368},
  \doiprefix\url{10.1016/j.neuropharm.2007.05.018} (\bibinfo{year}{2007}).

\bibitem{Wang2014}
\bibinfo{author}{Wang, J.~Q.} \emph{et~al.}
\newblock \bibinfo{journal}{\bibinfo{title}{Roles of subunit phosphorylation in
  regulating glutamate receptor function}}.
\newblock {\emph{\JournalTitle{European Journal of Pharmacology}}}
  \textbf{\bibinfo{volume}{728}}, \bibinfo{pages}{183--187},
  \doiprefix\url{10.1016/j.ejphar.2013.11.019} (\bibinfo{year}{2014}).

\bibitem{LBuchachenko2014}
\bibinfo{author}{Buchachenko, A.~L.}
\newblock \bibinfo{journal}{\bibinfo{title}{Magnetic control of enzymatic
  phosphorylation}}.
\newblock {\emph{\JournalTitle{Journal of Physical Chemistry \& Biophysics}}}
  \textbf{\bibinfo{volume}{2}}, \doiprefix\url{10.4172/2161-0398.1000142}
  (\bibinfo{year}{2014}).

\bibitem{Efimova2008}
\bibinfo{author}{Efimova, O.} \& \bibinfo{author}{Hore, P.}
\newblock \bibinfo{journal}{\bibinfo{title}{Role of exchange and dipolar
  interactions in the radical pair model of the avian magnetic compass}}.
\newblock {\emph{\JournalTitle{Biophysical Journal}}}
  \textbf{\bibinfo{volume}{94}}, \bibinfo{pages}{1565--1574},
  \doiprefix\url{10.1529/biophysj.107.119362} (\bibinfo{year}{2008}).

\bibitem{Hore2019}
\bibinfo{author}{Hore, P.}
\newblock \bibinfo{journal}{\bibinfo{title}{Upper bound on the biological
  effects of 50/60 hz magnetic fields mediated by radical pairs}}.
\newblock {\emph{\JournalTitle{{eLife}}}} \textbf{\bibinfo{volume}{8}},
  \doiprefix\url{10.7554/elife.44179} (\bibinfo{year}{2019}).

\bibitem{Kattnig2016}
\bibinfo{author}{Kattnig, D.~R.} \emph{et~al.}
\newblock \bibinfo{journal}{\bibinfo{title}{Chemical amplification of magnetic
  field effects relevant to avian magnetoreception}}.
\newblock {\emph{\JournalTitle{Nature Chemistry}}}
  \textbf{\bibinfo{volume}{8}}, \bibinfo{pages}{384--391},
  \doiprefix\url{10.1038/nchem.2447} (\bibinfo{year}{2016}).

\bibitem{Kattnig2017}
\bibinfo{author}{Kattnig, D.~R.} \& \bibinfo{author}{Hore, P.~J.}
\newblock \bibinfo{journal}{\bibinfo{title}{The sensitivity of a radical pair
  compass magnetoreceptor can be significantly amplified by radical
  scavengers}}.
\newblock {\emph{\JournalTitle{Scientific Reports}}}
  \textbf{\bibinfo{volume}{7}}, \doiprefix\url{10.1038/s41598-017-09914-7}
  (\bibinfo{year}{2017}).

\bibitem{Player2021}
\bibinfo{author}{Player, T.~C.}, \bibinfo{author}{Baxter, E. D.~A.},
  \bibinfo{author}{Allatt, S.} \& \bibinfo{author}{Hore, P.~J.}
\newblock \bibinfo{journal}{\bibinfo{title}{Amplification of weak magnetic
  field effects on oscillating reactions}}.
\newblock {\emph{\JournalTitle{Scientific Reports}}}
  \textbf{\bibinfo{volume}{11}}, \doiprefix\url{10.1038/s41598-021-88871-8}
  (\bibinfo{year}{2021}).

\bibitem{Binhi2023}
\bibinfo{author}{Binhi, V.~N.}
\newblock \bibinfo{journal}{\bibinfo{title}{Statistical amplification of the
  effects of weak magnetic fields in cellular translation}}.
\newblock {\emph{\JournalTitle{Cells}}} \textbf{\bibinfo{volume}{12}},
  \bibinfo{pages}{724}, \doiprefix\url{10.3390/cells12050724}
  (\bibinfo{year}{2023}).

\bibitem{Forbes2007}
\bibinfo{author}{Forbes, M. D.~E.} \emph{et~al.}
\newblock \bibinfo{journal}{\bibinfo{title}{On the electron spin polarization
  observed in {TREPR} experiments involving hydroxyl and sulfate radicals}}.
\newblock {\emph{\JournalTitle{Molecular Physics}}}
  \textbf{\bibinfo{volume}{105}}, \bibinfo{pages}{2127--2136},
  \doiprefix\url{10.1080/00268970701663768} (\bibinfo{year}{2007}).

\bibitem{Heo2006}
\bibinfo{author}{Heo, J.} \& \bibinfo{author}{Campbell, S.~L.}
\newblock \bibinfo{journal}{\bibinfo{title}{Ras regulation by reactive oxygen
  and nitrogen species}}.
\newblock {\emph{\JournalTitle{Biochemistry}}} \textbf{\bibinfo{volume}{45}},
  \bibinfo{pages}{2200--2210}, \doiprefix\url{10.1021/bi051872m}
  (\bibinfo{year}{2006}).

\bibitem{Liming1969}
\bibinfo{author}{Liming, F.~G.}
\newblock \bibinfo{journal}{\bibinfo{title}{Free radicals formed in aliphatic
  polyamino acids by exposure to hydrogen atoms}}.
\newblock {\emph{\JournalTitle{Radiation Research}}}
  \textbf{\bibinfo{volume}{39}}, \bibinfo{pages}{252},
  \doiprefix\url{10.2307/3572665} (\bibinfo{year}{1969}).

\bibitem{Neese2012}
\bibinfo{author}{Neese, F.}
\newblock \bibinfo{journal}{\bibinfo{title}{The {ORCA} program system}}.
\newblock {\emph{\JournalTitle{{WIREs} Computational Molecular Science}}}
  \textbf{\bibinfo{volume}{2}}, \bibinfo{pages}{73--78},
  \doiprefix\url{10.1002/wcms.81} (\bibinfo{year}{2012}).

\end{thebibliography}

% \noindent LaTeX formats citations and references automatically using the bibliography records in your .bib file, which you can edit via the project menu. Use the cite command for an inline citation, e.g.  \cite{Hao:gidmaps:2014}.

% For data citations of datasets uploaded to e.g. \emph{figshare}, please use the \verb|howpublished| option in the bib entry to specify the platform and the link, as in the \verb|Hao:gidmaps:2014| example in the sample bibliography file.

\section*{Acknowledgements}

The authors would like to thank Dennis Salahub for helpful feedback. This work was supported by the Natural Sciences and Engineering Research Council (NSERC) and the National Research Council (NRC) of Canada. 

\section*{Author contributions statement}

H.ZH. and C.S. conceived the project; P.S.N. conducted the investigation and performed the modelling \& calculations with help from H.ZH. and C.S.; P.S.N. wrote the paper with feedback from H.ZH. and C.S. 

\section*{Competing Interests}

The authors declare no competing interests.

\section*{Appendix}
\setcounter{figure}{0}

\makeatletter 
\renewcommand{\thefigure}{\@Roman\c@figure}
\makeatother

\subsection*{A. Magnesium Isotope Effect}

We performed calculations of FSY for the RPs with the spinful and spinless isotopes taken separately to understand the potential effects of the spin on the behaviour of the system. 

\subsubsection*{Ser Pathway}

\begin{figure}[h!]
\begin{subfigure}{.49\textwidth}
  \centering
  % include first image
\includegraphics[width=0.97\linewidth]{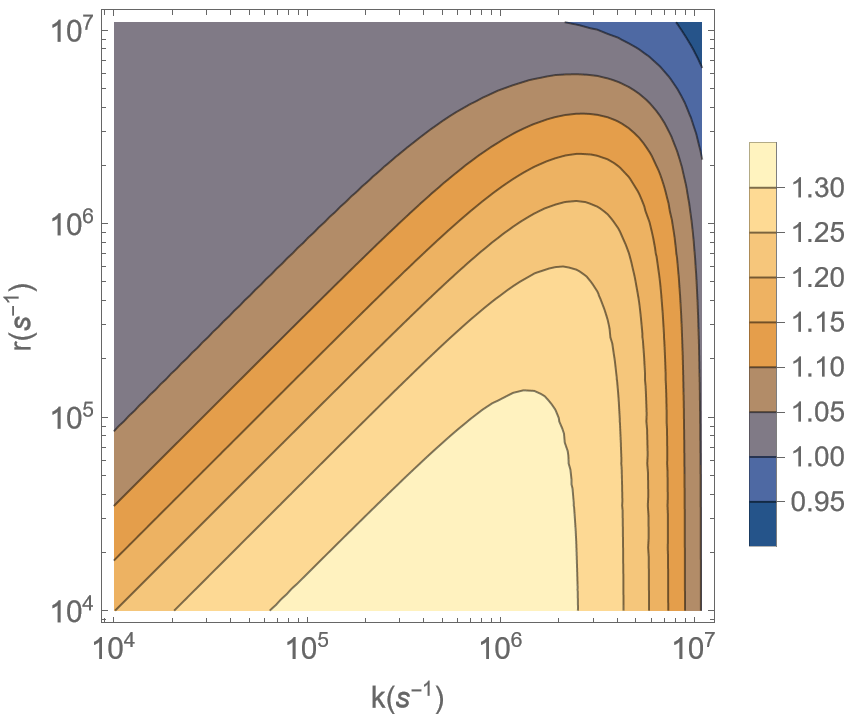}
  \caption{}
  \label{subfig:1.1}
\end{subfigure}
\begin{subfigure}{.481\textwidth}
  \centering
  % include second image
  \includegraphics[width=0.97\linewidth]{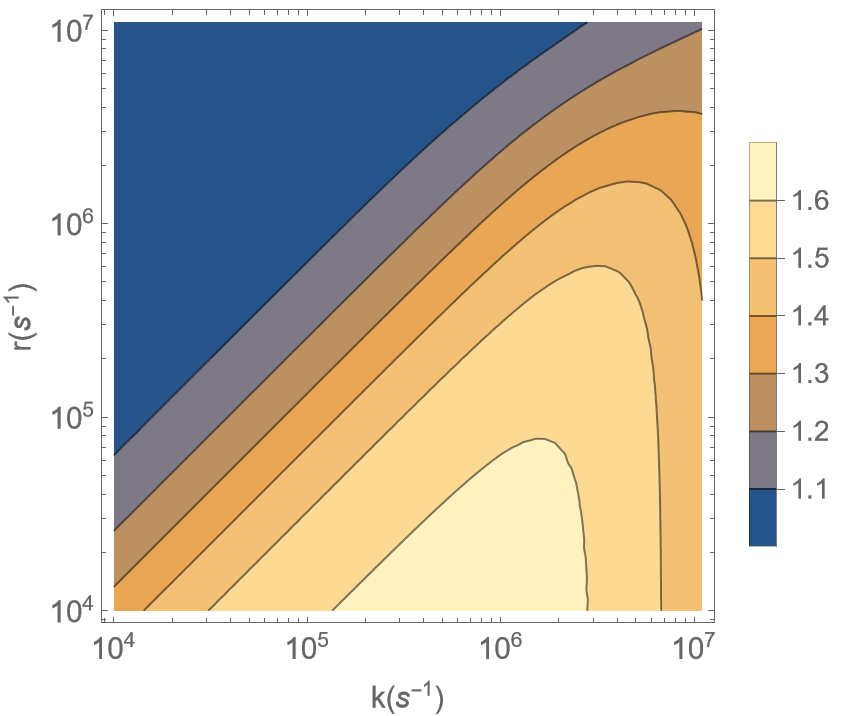}  
  \caption{}
  \label{subfig:1.2}
\end{subfigure}
\caption{Ratio of the fractional singlet yield percentages, $S$ plotted against $k(\mbox{s}^{-1})$ and $r(\mbox{s}^{-1})$ for Ser radical case: (a) System containing only $^{24}$Mg and $^{26}$Mg isotopes is considered; (b) System containing $^{25}$Mg isotope with the HFCC, $a_B = -11.22$ mT has been considered.}
\label{subfig:1}
\end{figure}

Fig.\ref{subfig:1} shows the quantity $S$ for the radical pair with the serine oxyradical plotted against the reaction rate, $k$, and relaxation rate, $r$. Both calculations consider the contribution from the nucleus of Serine oxyradical that has the highest HFCC, . Fig.\ref{subfig:1.1} shows the behaviour when the calculations consider the spinless isotopes, $^{24}$Mg and $^{26}$Mg. Fig.\ref{subfig:1.2} shows the case where the HFI contribution from the spinful isotope, $^{25}$Mg, is considered. We find that higher values of $S$ are found when only the spinful isotope is considered for the calculation.

\begin{figure}[h!]
\begin{subfigure}{.481\textwidth}
  \centering
  % include first image
  \includegraphics[width=0.95\linewidth]{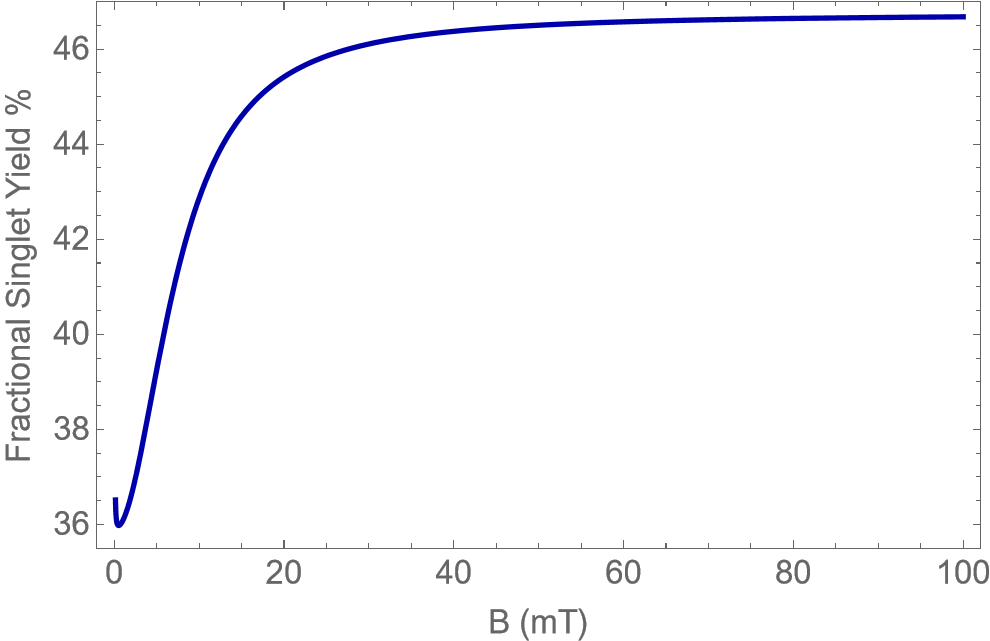}
  \caption{}
  \label{subfig:2.1}
\end{subfigure}
\begin{subfigure}{.49\textwidth}
  \centering
  % include second image
  \includegraphics[width=0.95\linewidth]{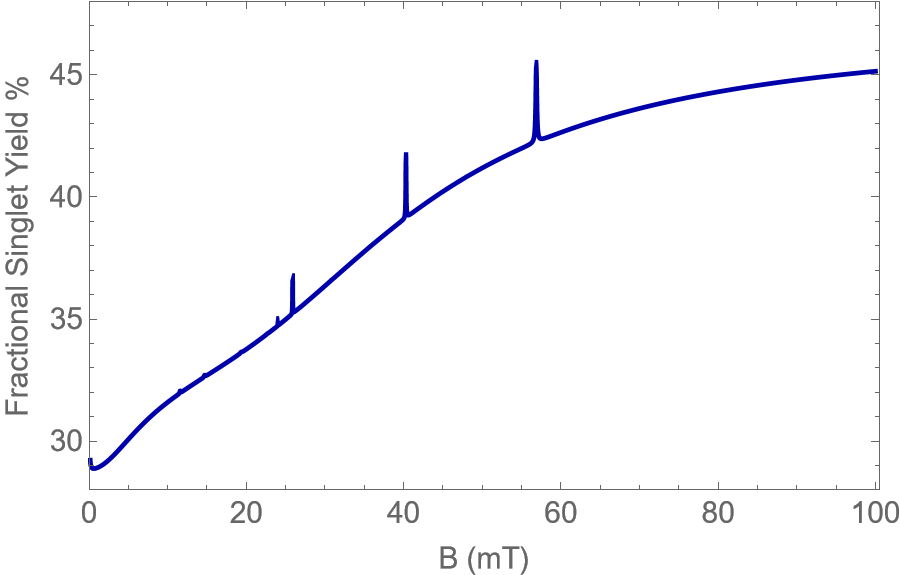}  
  \caption{}
  \label{subfig:2.2}
\end{subfigure}
\caption{Variation of the fractional singlet yield percentage with magnetic field strength ranging from $0.15$ mT to $100$ mT for the Tyr Pathway. The figure is plotted for the choice of $k =  2\times 10^6\mbox{ s}^{-1}$ and $r = 3 \times10^5\mbox{ s}^{-1}$: (a) System containing only $^{24}$Mg and $^{26}$Mg isotopes is considered; (b) System containing $^{25}$Mg isotope with the HFCC, $a_B = -11.22$ mT has been considered.}
\label{subfig:2}
\end{figure}

An example of the variation of fractional singlet yield percentage with magnetic field strength from $0.15\mbox{ mT}$ to $100\mbox{ mT}$ for a choice of $k$ and $r$ from the feasible region is given in Figs. \ref{subfig:2.1} and \ref{subfig:2.2}.  We choose $k =  2\times 10^6\mbox{ s}^{-1}$ and $r = 3 \times10^5\mbox{ s}^{-1}$ for the same. We find that the behaviour of the system is significantly different between the spinless and spinful cases. The singlet yield at low field are substantially lower for the spinful case. There are sudden spikes in the value at certain MF strengths, as well, which can be associated with ethe igenenergies of certain states crossing over.

\subsubsection*{Tyr Pathway}
\begin{figure}[h!]
\begin{subfigure}{.49\textwidth}
  \centering
  % include first image
  \includegraphics[width=1\linewidth]{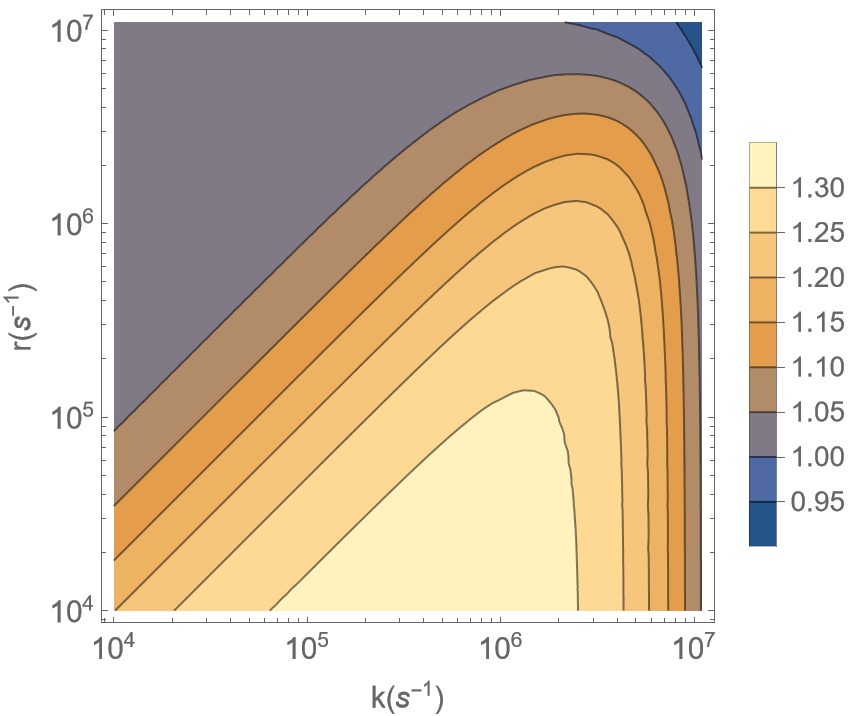}
  \caption{}
  \label{subfig:3.1}
\end{subfigure}
\begin{subfigure}{.481\textwidth}
  \centering
  % include second image
  \includegraphics[width=1\linewidth]{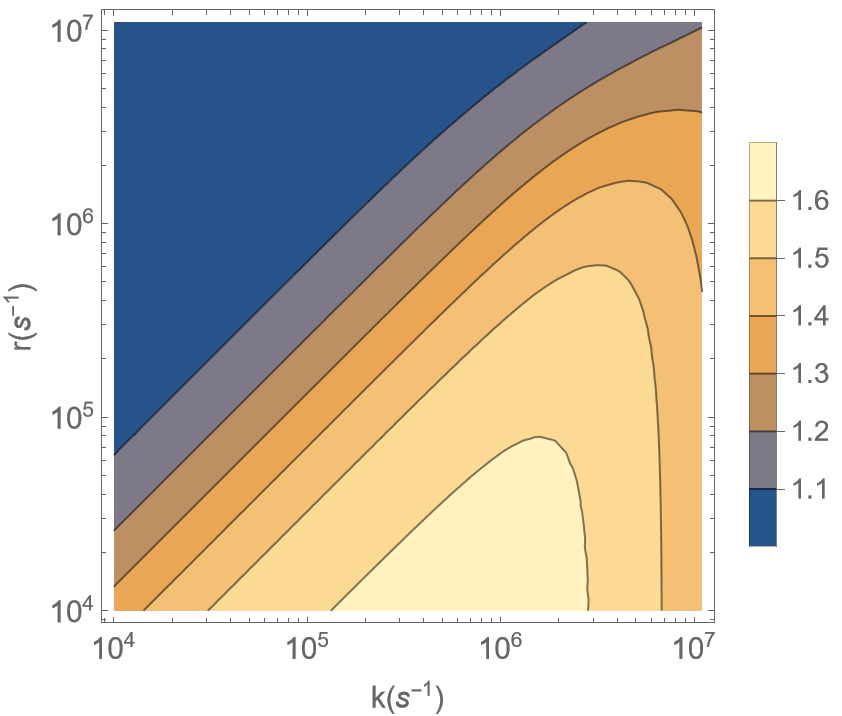}
  \caption{}
  \label{subfig:3.2}
\end{subfigure}
\caption{Ratio of the fractional singlet yield percentages, $S$ plotted against $k(\mbox{s}^{-1})$ and $r(\mbox{s}^{-1})$ for Tyr radical case: (a) System containing only $^{24}$Mg and $^{26}$Mg isotopes is considered; (b) System containing $^{25}$Mg isotope with the HFCC, $a_B = -11.22$ mT has been considered.}
\label{subfig:3}
\end{figure}

Similar plots of the quantity $S$ as in the previous section were plotted for the radical pair involving the tyrosine oxyradical in Fig. \ref{subfig:3}. We consider the HFCC contribution from the hydrogen nucleus of the Tyrosine radical having the highest HFCC, $a_A = 1.86$ mT, for all the plots. 

\begin{figure}[h!]
\begin{subfigure}{.481\textwidth}
  \centering
  % include first image
  \includegraphics[width=1\linewidth]{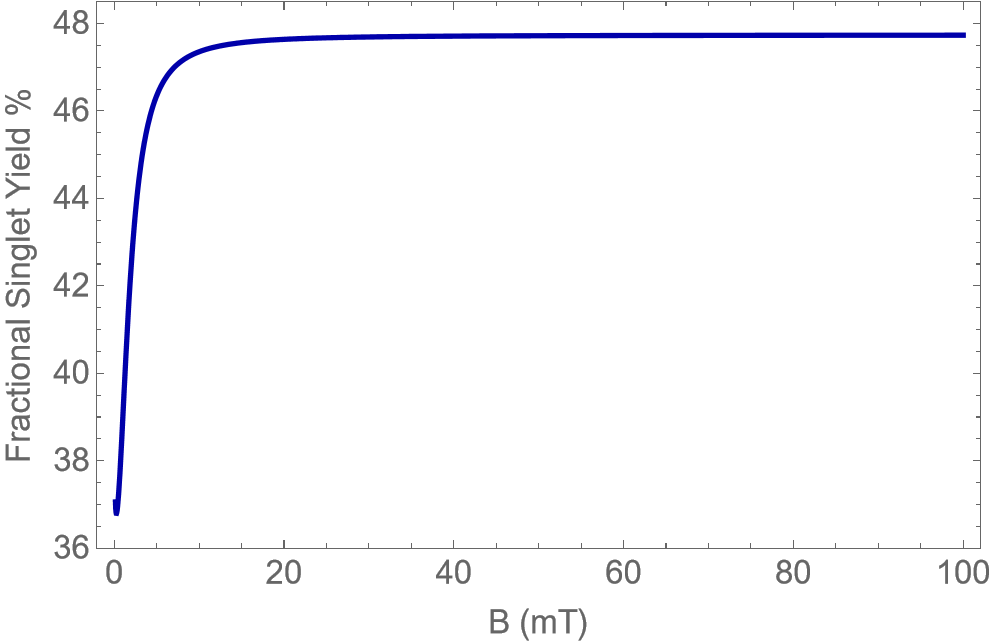}
  \caption{}
  \label{subfig:4.1}
\end{subfigure}
\begin{subfigure}{.49\textwidth}
  \centering
  % include second image
  \includegraphics[width=1\linewidth]{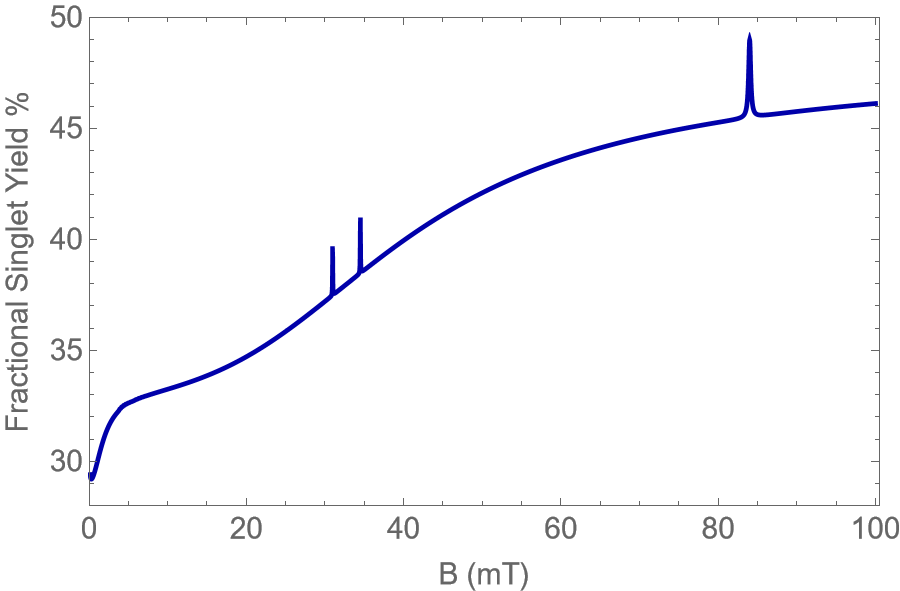}  
  \caption{}
  \label{subfig:4.2}
\end{subfigure}
\caption{Variation of the fractional singlet yield percentage with magnetic field strength ranging from $0.15$ mT to $100$ mT for the Tyr Pathway. The figure is plotted for the choice of $k =  2\times 10^6\mbox{ s}^{-1}$ and $r = 2 \times10^5\mbox{ s}^{-1}$: (a) System containing only $^{24}$Mg and $^{26}$Mg isotopes is considered; (b) System containing $^{25}$Mg isotope with the HFCC, $a_B = -11.22$ mT has been considered.}
\label{subfig:4}
\end{figure}

Figs. \ref{subfig:4.1} and \ref{subfig:4.2} show examples of the variation of fractional singlet yield percentage with magnetic field strength from $0.15\mbox{ mT}$ to $100\mbox{ mT}$ for the choice of $k =  2\times 10^6\mbox{ s}^{-1}$ and $r = 2 \times10^5\mbox{ s}^{-1}$ which fall in the range of possible values of $k$ and $r$.

The differences that one observes between the spinful and spinless isotopes for this case are similar to those for the Ser pathway.

\subsection*{B. Isotope Contributions at Low Field}

\begin{figure}[h!]
\begin{subfigure}{.49\textwidth}
  \centering
  % include first image
  \includegraphics[width=1\linewidth]{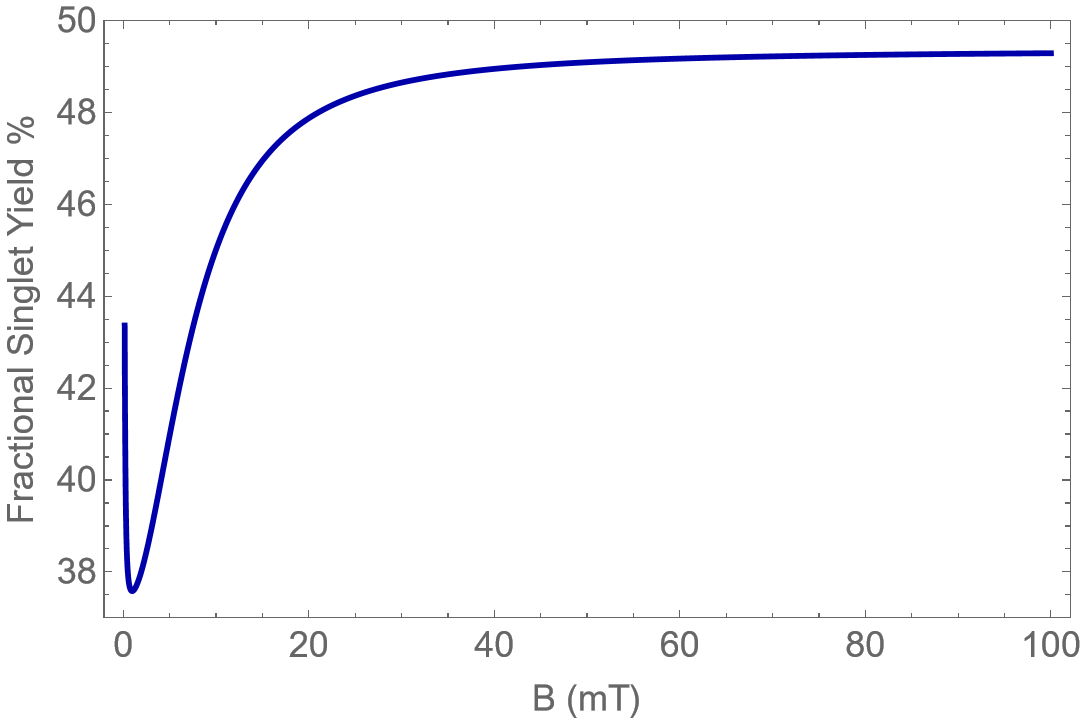}
  \caption{Ser Pathway}
  \label{fig:7.1}
\end{subfigure}
\begin{subfigure}{.49\textwidth}
  \centering
  % include second image
  \includegraphics[width=1\linewidth]{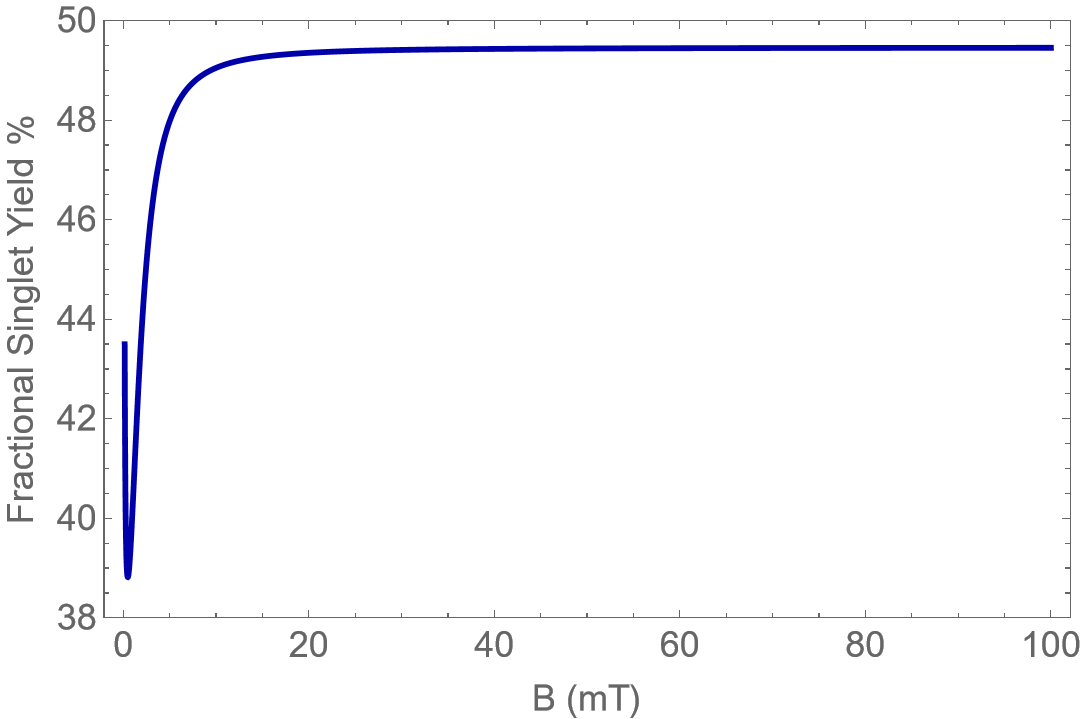}  
  \caption{Tyr Pathway}
  \label{fig:7.2}
\end{subfigure}

\begin{subfigure}{.49\textwidth}
  \centering
  % include second image
  \includegraphics[width=1\linewidth]{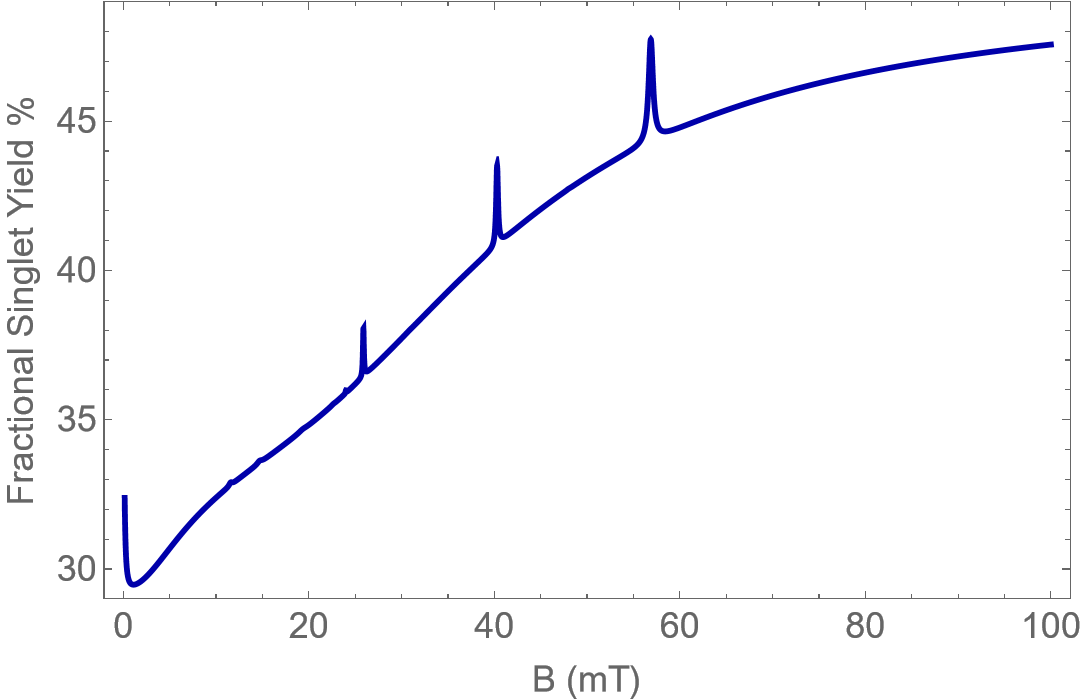}  
  \caption{Ser Pathway}
  \label{fig:7.3}
\end{subfigure}
\begin{subfigure}{.49\textwidth}
  \centering
  % include second image
  \includegraphics[width=1\linewidth]{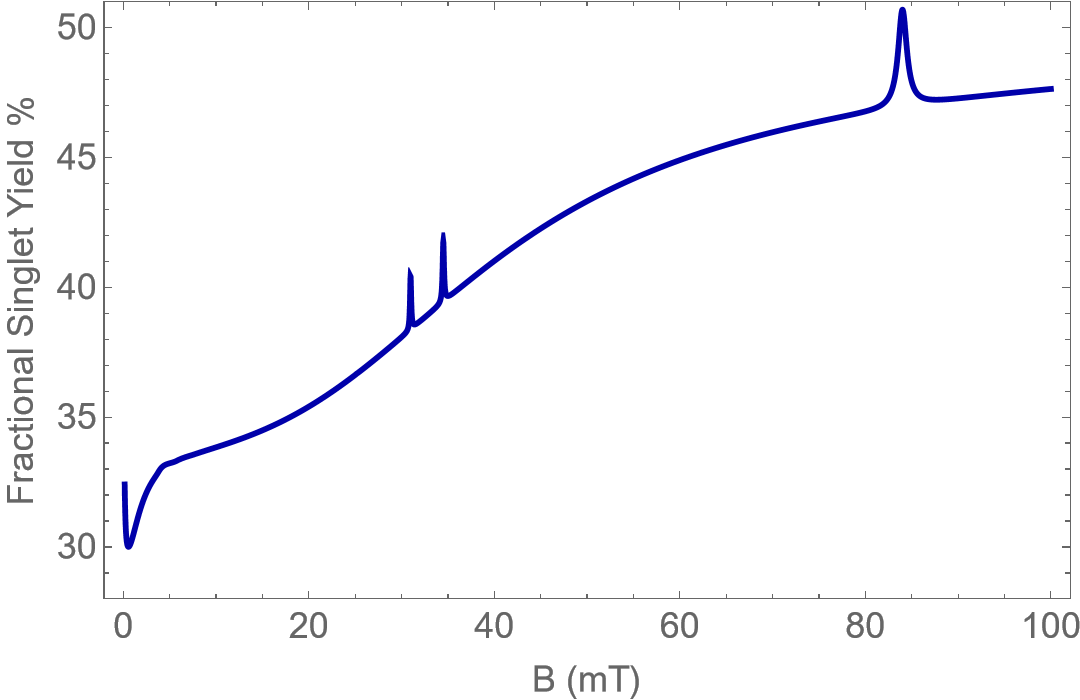}  
  \caption{Tyr Pathway}
  \label{fig:7.4}
\end{subfigure}
\caption{Variation of the FSY percentage with magnetic field strength ranging from $0.15$ mT to $100$ mT plotted with $k =  7.5\times 10^6\mbox{ s}^{-1}$ and $r = 2\times10^5\mbox{ s}^{-1}$. (a) and (b) are plotted considering a system with only $^{24}$Mg and $^{26}$Mg isotopes; (c) and (d) are plotted considering system with only $^{25}$Mg isotope at its natural abundance of 10\%. The HFI contribution from the same is $a_B = -11.22$ mT.}
\label{fig:7}
\end{figure}

We plot examples in Fig.\ref{fig:7} from both the Ser and Tyr pathway by considering the contribution from the spinless and spinful isotopes separately to understand the contribution of the same to the drop in FSY at MF strength close to 0 mT. We find that the drop in FSY values occur with or without the contribution from the $^{25}$Mg isotope.

\subsection*{C. HFCCs of Other Nuclei}

For the calculations in this paper, we consider only the HFI contribution of the nuclei of the oxyradical with the highest HFCC. This is a common approximation that is used in such calculations for the sake of simplicity. 

The next highest HFCC of a nucleus in the Serine oxyradical was 5.73 mT. Similarly, in the case of the Tyrosine oxyradical, the next highest HFCCs were -0.64 mT and -0.53 mT.

\end{document}